\documentclass{jpp}

\usepackage[dvipdfmx]{graphicx}
\usepackage[dvipdfmx]{color}
\usepackage{natbib}
\usepackage{bm}
\usepackage{amsmath}

\newcommand{\note}[1]{\textcolor{black}{#1}}
\newcommand{\nuno}[1]{\textcolor{black}{#1}}
\newcommand{\revise}[1]{\textcolor{black}{#1}}

\newcommand{\EB}{\bm{E} \times \bm{B}}

\newcommand{\diff}{\mathrm{d}}

\newcommand{\nuii}{\nu_{\mathrm{ii}}}
\newcommand{\nuee}{\nu_{\mathrm{ee}}}

\newcommand{\nuei}{\nu_{\mathrm{ei}}}

\newcommand{\nue}{\nu_{\mathrm{e}}}

\newcommand{\vthi}{v_{\mathrm{th,i}}}
\newcommand{\vthe}{v_{\mathrm{th,e}}}

\newcommand{\mi}{m_{\mathrm{i}}}
\newcommand{\me}{m_{\mathrm{e}}}
\newcommand{\Ti}{T_{0\mathrm{i}}}
\newcommand{\Te}{T_{0\mathrm{e}}}
\newcommand{\rhoi}{\rho_{\mathrm{i}}}
\newcommand{\rhoe}{\rho_{\mathrm{e}}}
\newcommand{\rhos}{\rho_{\mathrm{Se}}}
\newcommand{\de}{d_{\mathrm{e}}}
\newcommand{\di}{d_{\mathrm{i}}}

\newcommand{\tauA}{\tau_{\mathrm{A}}}
\newcommand{\VA}{V_{\mathrm{A}}}
\newcommand{\betae}{\beta_{\mathrm{e}}}
\newcommand{\betai}{\beta_{\mathrm{i}}}

\ifCUPmtlplainloaded \else
  \checkfont{eurm10}
  \iffontfound
    \IfFileExists{upmath.sty}
      {\typeout{^^JFound AMS Euler Roman fonts on the system,
                   using the 'upmath' package.^^J}%
       \usepackage{upmath}}
      {\typeout{^^JFound AMS Euler Roman fonts on the system, but you
                   dont seem to have the}%
       \typeout{'upmath' package installed. JPP.cls can take advantage
                 of these fonts, if you use 'upmath' package.^^J}%
      }
  \else
  \fi
\fi

\ifCUPmtlplainloaded \else
  \checkfont{msam10}
  \iffontfound
    \IfFileExists{amssymb.sty}
      {\typeout{^^JFound AMS Symbol fonts on the system, using the
                'amssymb' package.^^J}%
       \usepackage{amssymb}%
         \let\leq=\leqslant
         \let\geq=\geqslant
      }{}
  \fi
\fi

\ifCUPmtlplainloaded \else
  \IfFileExists{amsbsy.sty}
    {\typeout{^^JFound the 'amsbsy' package on the system, using it.^^J}%
     \usepackage{amsbsy}}
    {}
\fi

\newsavebox{\astrutbox}
\sbox{\astrutbox}{\rule[-5pt]{0pt}{20pt}}

\newcommand\p{\ensuremath{\partial}}

\newcommand\eg{e.g.\ }
\newcommand\ie{i.e.\ }

\title[Ion and electron heating during magnetic reconnection]{Ion and
electron heating during magnetic reconnection in weakly collisional
plasmas}

\author[R. Numata and N. F. Loureiro]%
{R\ls Y\ls U\ls S\ls U\ls K\ls E\ns N\ls U\ls M\ls A\ls T\ls A$^1$%
  \thanks{Email address for correspondence: ryusuke.numata@gmail.com}\ns
\and N.\ns F.\ns L\ls O\ls U\ls R\ls E\ls I\ls R\ls O$^2$}

\affiliation{$^1$Graduate School of Simulation Studies, University of Hyogo,
7-1-28 Minatojima Minami-machi, Chuo-ku, Kobe, Hyogo 650-0047, Japan\\[\affilskip]
$^2$Instituto de Plasmas e Fus\~ao
Nuclear,
Instituto Superior T\'ecnico,
Universidade de Lisboa, 1049-001 Lisboa, Portugal}

\pubyear{2010}
\volume{650}
\pagerange{119--126}
\date{24 June 2014; revised 21 October 2014; accepted 21
October 2014; first publised online 28 November 2014}
\begin{document}

\maketitle

\begin{abstract}
 Magnetic reconnection and associated heating of ions and electrons in
 strongly magnetized, weakly collisional plasmas are studied by means of
 gyrokinetic simulations. It is shown that an appreciable amount of the
 released magnetic energy is dissipated to yield (irreversible) electron
 and ion heating via phase mixing.
 Electron heating is \nuno{mostly} localized to the magnetic island, not the
 current sheet, and occurs after the dynamical reconnection stage.
 Ion heating is comparable to electron
 heating only in high-$\beta$ plasmas, and results from
 both parallel and perpendicular phase mixing
 due to finite Larmor radius (FLR) effects; in space, ion heating is mostly
 localized to the interior of a secondary island (plasmoid) that arises
 from the instability of the current sheet.
\end{abstract}

 \begin{PACS}
  52.35.Vd, 52.35.Py, 52.30.Gz, 52.65.Tt
 \end{PACS}

\section{Introduction}
\label{sec:intro}

Magnetic reconnection is a commonly observed fundamental process in
plasmas. It allows topological change of magnetic field lines, and
rapidly converts the free energy stored in the magnetic field into
various forms of energy.
Amongst other poorly understood aspects of magnetic reconnection, such
as explaining the explosive time scale that is often observed, or the
onset mechanism, one key issue is the question of energy partition, \ie
understanding how, and how much of, the released energy is distributed
into the different available channels: bulk heating of each of the
plasma species and non-thermal particle acceleration.

Plasma heating is often observed to accompany magnetic reconnection in
both astrophysical and laboratory plasmas (see, \eg a recent review
by \citet{YamadaKulsrudJi_10}).
Specifically, the measured ion temperature in reversed-field pinches (RFP),
spheromaks, and merging plasma
experiments where reconnection events are thought to be
occurring~\citep{OnoYamadaAkao_96,HsuCarterFiksel_01,FikselAlmagriChapman_09}
often well exceeds the electron temperature. The fact that ions are
selectively heated invalidates Ohmic dissipation of the current sheet as
the dominant heating source since it \note{primarily} deposits the
energy into current carrying electrons. Furthermore, heating associated
with reconnection events generally occurs much faster than the expected
time scale estimated from the Ohmic heating, as expected in weakly
collisional environments.
Various mechanisms causing such `anomalous' ion heating have been
suggested: Stochastic ion heating has been
studied~\citep{FikselAlmagriChapman_09} for explaining ion heating in a
RFP device where multiple reconnections in turbulent
environments are occurring rather than a single reconnection event,
\note{while anomalous resistivity has been invoked in MRX~\citep
{HsuCarterFiksel_01}.}

In the weakly collisional plasmas often found in reconnection environments, 
Ohmic and viscous heating cannot, by definition, be important. This leaves
phase mixing as the only possible mechanism of converting energy from 
the fields to the particles in an irreversible way.
Indeed, kinetic effects generally lead to
non-Maxwellian distribution functions; once highly oscillatory structures
are created in velocity space, they suffer strong collisional
dissipation as the collision operator provides diffusion in velocity
space. The thermalization time scale may be comparable to the time scale of
magnetic reconnection; therefore, to address thermodynamic properties of such
plasmas, collisional effects are essential, even though collisions are rare 
and not responsible for reconnection itself (\ie the frozen flux condition
is broken by electron kinetic effects, not collisions)
\footnote{In the interest of clarity, notice that by `heating' we mean 
the increase in the entropy of the system due to reconnection.
Note that this definition excludes the energization 
of supra-thermal particles, a process that is sometimes also 
referred to as heating
but which is formally reversible and does not, therefore, change 
the entropy of the system by itself.
}.


Landau damping~\citep{Landau_46} is one of the well-known
examples of phase mixing, in which nearly synchronous particles with the
electrostatic potential variations absorb energy from the field. Phase
mixing occurs along the 
direction of electron free stream, \ie along the magnetic field lines.
The parallel phase mixing and resultant heating of electrons during
magnetic reconnection in low-$\beta$ plasmas ($\beta$ is the ratio of the
plasma to the magnetic pressure) have been studied using a reduced
kinetic model~\citep{ZoccoSchekochihin_11,LoureiroSchekochihinZocco_13}
and the full gyrokinetic model~\citep{NumataLoureiro_14}, and the
significance of such an effect has been proved.

Phase mixing can also be induced by a FLR
effect~\citep{DorlandHammett_93}. In strongly magnetized plasmas,
particles undergo drifts {(dominantly the $\bm{E}\times\bm{B}$ drift) in
the perpendicular direction to the mean
magnetic field. Gyro-averaging of the fields will give rise to
spread of the drift velocity of different particles, hence will lead
to phase mixing in the perpendicular 
direction. Unlike linear parallel phase mixing of Landau damping,
the FLR-phase-mixing process causes damping proportional to the field
strength, and only appears nonlinearly (nonlinear phase mixing).
Energy dissipation due to FLR phase mixing has 
been investigated in electrostatic 
turbulence~\citep{TatsunoDorlandSchekochihin_09}; its role in magnetic
reconnection has never been studied.

In this paper, we present gyrokinetic simulations of magnetic
reconnection using the numerical code {\tt
AstroGK}~\citep{NumataHowesTatsuno_10}.
Our main aim is to perform a detailed analysis of ion and electron
(irreversible) heating in weakly collisional, strongly magnetized
plasmas.
We follow \citet{NumataDorlandHowes_11} and \citet{NumataLoureiro_14} to
setup a tearing-mode unstable configuration.
In addition to the electron heating
due to parallel phase mixing in low-$\beta$ plasmas as has already been
shown in \citet{LoureiroSchekochihinZocco_13}
and \citet{NumataLoureiro_14}, ion heating 
is also expected to be significant in high-$\beta$ plasmas since
compressible fluctuations will be excited which are strongly damped
collisionlessly~\citep{SchekochihinCowleyDorland_09}.
The main questions we are going to address will be (i) fraction of ion
and electron heating, (ii) dissipation channel, \ie how much energy is
converted via phase mixing, resistivity and viscosity,
and (iii) how the answer to the above questions scales with plasma beta.
We also study in detail how and where plasma
heating occurs via phase mixing, which may be important to the
interpretation of plasma heating as measured in laboratory experiments
or space and astrophysical plasmas.

This paper is organized as follows:
We start with defining plasma heating and energy dissipation within the
framework of the gyrokinetic model in Sec.~\ref{sec:heating}.
The details of the simulation setup and choices of parameters are
described in Sec.~\ref{sec:setup}. In the same section, simulations of
the linear regime are presented to identify the region of parameter
space where the tearing instability growth rate is independent of the
plasma collisionality (the so-called collisionless regime of reconnection).
In Sec.~\ref{sec:results}, we report the results of nonlinear
simulations, focusing on the dependence on plasma beta of the
reconnection dynamics, as well as on the detailed analysis of ion and
electron heating caused by magnetic reconnection. We summarize the main
results of this paper in Sec.~\ref{sec:conclusion}.

\section{Plasma heating and energy dissipation}
\label{sec:heating}

We consider magnetic reconnection in the framework of $\delta f$
gyrokinetics. 
Plasma heating is measured by the entropy production (irreversible),
rather than increase of temperature (reversible) in this work.
In the absence of collisions, the gyrokinetic equation
conserves the generalized energy consisting of the particle part
$E^{\mathrm{p}}_{s}$ and the magnetic field part
$E^{\mathrm{m}}_{\perp,\parallel}$ 
\begin{equation}
 W = \sum_{s} E^{\mathrm{p}}_s + E^{\mathrm{m}}_{\perp} +
  E^{\mathrm{m}}_{\parallel} = 
  \int
  \left[
   \sum_{s}\int \frac{T_{0s} \delta f_{s}^{2}}{2f_{0s}} \diff \bm{v}
   + \frac{\left|\nabla_{\perp}A_{\parallel}\right|^{2}}{2\mu_{0}}
   + \frac{\left|\delta B_{\parallel}\right|^{2}}{2\mu_{0}}
  \right] \diff \bm{r}.
  \label{eq:gen_energy}
\end{equation} 
The subscript $s=\mathrm{i}$ (for ions) or $\mathrm{e}$ (for
 electrons) denotes the species label.
 $f_{0s}=n_{0s}/(\sqrt{\pi}v_{\mathrm{th},s})^{3}
 \exp(-v^{2}/v_{\mathrm{th},s}^{2})$ is the Maxwellian distribution function
 of background plasmas, $n_{0s}$, $T_{0s}$ are the density and
 temperature, $v_{\mathrm{th},s}\equiv\sqrt{2T_{0s}/m_{s}}$ is the
 thermal velocity, $m_{s}$, $q_{s}$ are the mass and  electric
 charge. 
 The perturbed distribution function \nuno{is} $\delta f_{s}= -
 \left(q_{s}\phi/T_{0s}\right) f_{0s} + h_{s}$, where $h_{s}$ is the
 non-Boltzmann part obeying the gyrokinetic equation.
 $\phi$, $A_{\parallel}$, \nuno{$\delta B_{\parallel}$ are, respectively, the electrostatic
 potential, vector potential and perturbed magnetic field} along the background
 magnetic field $\bm{B_{0}}=B_{z0}\hat{z}$, and $\mu_{0}$ is the
 vacuum permeability.
 \nuno{Note that the perturbed particle energy is
 $E_{s}^{\mathrm{p}}=-T_{0s} \delta S_{s}$,
 where $\delta S_{s}$ is the perturbed entropy.}}

 \revise{
 If we 
 extract explicitly the first two velocity moments from $\delta f_{s}$ 
 as in \citet{ZoccoSchekochihin_11},
 \begin{equation}
  \delta f_{s} = \left(\frac{n_{s}}{n_{0s}} + \frac{2 v_{\parallel}
		  u_{\parallel,s}}{v_{\mathrm{th},s}^{2}}\right) f_{0s}
  + h_{s}',
  \label{eq:hermite}
 \end{equation}
 where
 \begin{equation}
  n_{s} = \int \delta f_s \diff \bm{v}, ~~~
  u_{\parallel,s} = \frac{1}{n_{0s}} \int v_{\parallel} \delta f_s \diff \bm{v},
 \end{equation}
 the particle's energy is decomposed into the density variance, the
 parallel kinetic energy, and the rest as follows:
 \begin{equation}
  E^{\mathrm{p}}_{s}
   = \int
   \left[
    \frac{n_{0s}T_{0s}}{2} \frac{n_s^{2}}{n_{0s}^{2}}
    + \frac{m_{s}n_{0s}u_{\parallel,s}^{2}}{2}
    + \int \frac{T_{0s} h_{s}'^2}{2f_{0s}} \diff \bm{v}
   \right] \diff \bm{r}.
 \end{equation}
}

 The generalized energy is dissipated by collisions as $\diff W/\diff t =
 -\sum_{s} D_{s}$, where the dissipation rate of each species is given by
 \begin{equation}
  D_{s} =
  -\int \int
  \left\langle
   \frac{T_{0s}h_{s}}{f_{0s}} 
   \left(\frac{\p h_{s}}{\p t}\right)_{\mathrm{coll}}
  \right\rangle_{\bm{r}}
  \diff \bm{r} \diff \bm{v} \geq 0.
  \label{eq:dissipation_rate}
 \end{equation}
The angle bracket denotes the gyro-averaging operation.
\note{The collision term $(\p h_{s}/\p t)_{\mathrm{coll}}$ is modeled by the
linearized, and gyro-averaged, Landau collision
operator~\citep{AbelBarnesCowley_08,BarnesAbelDorland_09}. 
It consists of like-particle collisions of electrons and ions whose
frequencies are given by $\nuee$ and $\nuii$, and inter-species
collisions of electrons with ions given by $\nuei=Z_{\mathrm{eff}}\nuee$
($Z_{\mathrm{eff}}$ is the effective ion charge, and is set to unity in
this paper).
The ion-electron collisions are subdominant compared with the ion-ion
collisions. 
The electron-ion collisions reproduce Spitzer resistivity~\citep{SpitzerHarm_53}
for which the electron-ion collision frequency and the resistivity
($\eta$) are related by $\eta/\mu_{0}=0.380 \nuei \de^{2}$
where $\de\equiv\sqrt{\me/(\mu_{0}n_{0\mathrm{e}}q_{\mathrm{e}}^{2})}$
is the electron skin depth.}

The dissipation rate $D_s$ contains all possible dissipation channels.
Besides the dissipation that can be modeled in fluid models,
\eg the resistivity, $D_{s}$ also contains dissipation
of higher order velocity moments due to phase mixing depending on the
form of $h_s$.

In our problem setup explained in Sec.~\ref{sec:setup},
the initial energy is contained in the perpendicular magnetic field
energy and the electron parallel kinetic energy (\ie the reconnecting
magnetic field and the current associated with it).
During the magnetic reconnection process, the initial energy is
distributed into other forms of energy in a reversible way, and is only
irreversibly dissipated from the system through collisions. \note{There
is no direct dissipation channel of the magnetic field energy. The
resistive dissipation is the collisional decay of particle's kinetic energy
(current) supporting the magnetic field.}

The increased entropy is turned into heat, and the background
temperature increases on a time scale much longer than the time scale 
considered in the simulation~\citep{HowesCowleyDorland_06}.
In this sense, we identify the energy dissipation
\eqref{eq:dissipation_rate} and plasma heating. The background
temperature $T_{0\mathrm{i,e}}$ does not change during simulations.

\section{Simulation setup}
\label{sec:setup}

We consider magnetic reconnection of strongly magnetized 
plasmas in a two-dimensional doubly periodic slab domain.
Simulations are carried out using the gyrokinetic code {\tt
AstroGK}~\citep{NumataHowesTatsuno_10}.
We initialize the system by a tearing unstable magnetic field
configuration (see \citet{NumataHowesTatsuno_10,NumataDorlandHowes_11}
for details). The equilibrium magnetic field profile is
\begin{equation}
\bm B=B_{z0}\hat z+B_{y}^{\mathrm{eq}}(x)\hat y, \quad 
 \revise{B_{y}^{\mathrm{eq}}/B_{z0} \sim \varepsilon \ll 1},
\end{equation}
where $B_{z0}$ is the background guide magnetic field and
$B_{y}^{\mathrm{eq}}$ is the in-plane, reconnecting component, related
to the parallel vector potential by $B_{y}^{\mathrm{eq}}(x)= -\p
A_{\parallel}^{\mathrm{eq}}/\p x$, 
\revise{$\varepsilon$ is the gyrokinetic epsilon -- a small expansion
parameter enabling scale separation in gyrokinetics (see, \eg
\citet{HowesCowleyDorland_06})},
and
\begin{equation}
 \label{equilib_apar}
  A_{\parallel}^{\mathrm{eq}}(x) =
  A_{\parallel0}^{\mathrm{eq}}\cosh^{-2}\left(\frac{x-L_{x}/2}{a}\right)
  S_{\mathrm{h}}(x).
\end{equation}
($S_{\mathrm{h}}(x)$ is a shape function to enforce periodicity~\citep{NumataHowesTatsuno_10}.)
$A_{\parallel}^{\mathrm{eq}}$ is generated by the electron parallel
current to satisfy the parallel Amp\`ere's law.
The equilibrium scale length is denoted by $a$ and $L_x$ is the length
of the simulation box in the $x$ direction, set to $L_x/a=3.2\pi$. In
the $y$ direction, the box size is $L_y/a=2.5\pi$. We impose a small
sinusoidal perturbation to the equilibrium magnetic field,
$\tilde{A}_{\parallel} \propto \cos(k_y y)$ with wave number
$k_y a=2\pi a/L_y=0.8$, yielding a value of the tearing instability
parameter $\Delta'a\approx 23.2$.
The background temperatures ($T_{0\mathrm{i,e}}$) and densities
($n_{0\mathrm{i,e}}$) are spatially uniform and held constant throughout
the simulations.
We consider a quasi-neutral plasma, so
$n_{0\mathrm{i}}=n_{0\mathrm{e}}=n_{0}$, and singly charged ions
$q_\mathrm{i}=-q_{\mathrm{e}}=e$.

The equilibrium magnetic field defines the time scale of the system.
We normalize time by the Alfv\'en time $\tauA \equiv a/\VA$, where $\VA\equiv
 B_{y}^{\mathrm{max}}/\sqrt{\mu_{0}n_{0}\mi}$ is
 the (perpendicular) Alfv\'en velocity corresponding to the peak value of
 $B_{y}^{\mathrm{eq}}$.


We solve the fully electromagnetic gyrokinetic equations for
electrons and ions. {\tt AstroGK} employs a pseudo-spectral algorithm 
for the spatial coordinates ($x, y$), and Gaussian quadrature for
velocity space integrals. The velocity space is discretized in the
energy $E_s=m_s v^{2}/2$ and 
$\lambda=v_{\perp}^{2}/(B_{z0}v^{2})$. The velocity space derivatives in
the collision operator are estimated using a first order finite
difference scheme on an unequally spaced grid according to the
quadrature rules~\citep{BarnesAbelDorland_09}.

\subsection{Parameters}
\label{sec:param}

There are four basic parameters in the system:
The mass ratio, $\sigma\equiv \me/\mi$, the
temperature ratio of the background plasma, $\tau \equiv \Ti/\Te$, the
electron plasma beta, 
$\betae\equiv n_{0}\Te/(B_{z0}^{2}/2\mu_{0})$,
 and the ratio of the ion sound
 Larmor radius to the equilibrium scale length $a$, $\rhos/a\equiv 
c_{\mathrm{Se}}/(\Omega_{\mathrm{ci}}a)$. The ion sound speed for cold
 ions is
$c_{\mathrm{Se}}=\sqrt{\Te/\mi}$, and the ion cyclotron frequency is
$\Omega_{\mathrm{ci}}=e B_{z0}/\mi$. Those parameters define the
physical scales associated with the non-magnetohydrodynamic
effects: 
\begin{align}
 \rhoi = & \tau^{1/2} \rhos\sqrt{2}, &
 \di = & \betae^{-1/2} \rhos\sqrt{2}, \\
 \rhoe = & \sigma^{1/2} \rhos\sqrt{2}, &
 \de = & \betae^{-1/2} \sigma^{1/2} \rhos\sqrt{2}.
 \label{eq:ion_e_scales}
\end{align}
where $\rho_{\mathrm{i,e}}$ and $d_{\mathrm{i,e}}$ are the ion and
electron Larmor radii and skin depths, respectively.

We fix the following parameters throughout this paper:
\begin{align}
 \rhos/a = & 0.25/\sqrt{2}, &
 \sigma = & 0.01, &
 \tau = & 1.
\end{align}
Therefore, $\rhoi/a=0.25$, $\rhoe/a=0.025$, and $\betae=\betai$. We scan
in $\betae=\betai$ from 0.01 to 1; the electron inertial length also
changes accordingly as shown in Table~\ref{tbl:param}. The ion inertial
length is always $\di/\de=10$.
\begin{table}
 \begin{center}
  \begin{tabular}{cccccc}
   Case & $\betae$ & $\de/a$  & $\nue\tauA$\\ \\
   1 & 0.01 & 0.25 & $0.8 \times 10^{-4}$ \\
   2 & 0.03 & 0.14 & $1.4 \times 10^{-4}$ \\
   3 & 0.1 & 0.079 & $2.5 \times 10^{-4}$ \\
   4 & 0.3 & 0.046 & $4.4 \times 10^{-4}$ \\
   5 & 1 & 0.025 & $8.0 \times 10^{-4}$ \\
  \end{tabular}
  \caption{\label{tbl:param}Simulation parameters.}
 \end{center}
\end{table}

To identify the collisionless tearing-mode regime (\ie such that the
frozen-flux condition is not broken by collisions but instead by the
electron inertia or electron FLR effects),
we perform linear simulations.
For the linear runs, we take the number of collocation points in the
pitch angle direction ($\lambda$) and the energy direction ($E$) as
$N_\lambda =  N_{E}=16$; the number of grid points in the $x$
direction ranges from 256 to 1024, and we consider a single mode in the
$y$ direction (the fastest growing mode for this configuration).
We have verified that these numbers provide adequate resolution.

\begin{figure}
 \begin{center}
  \includegraphics[scale=0.45]{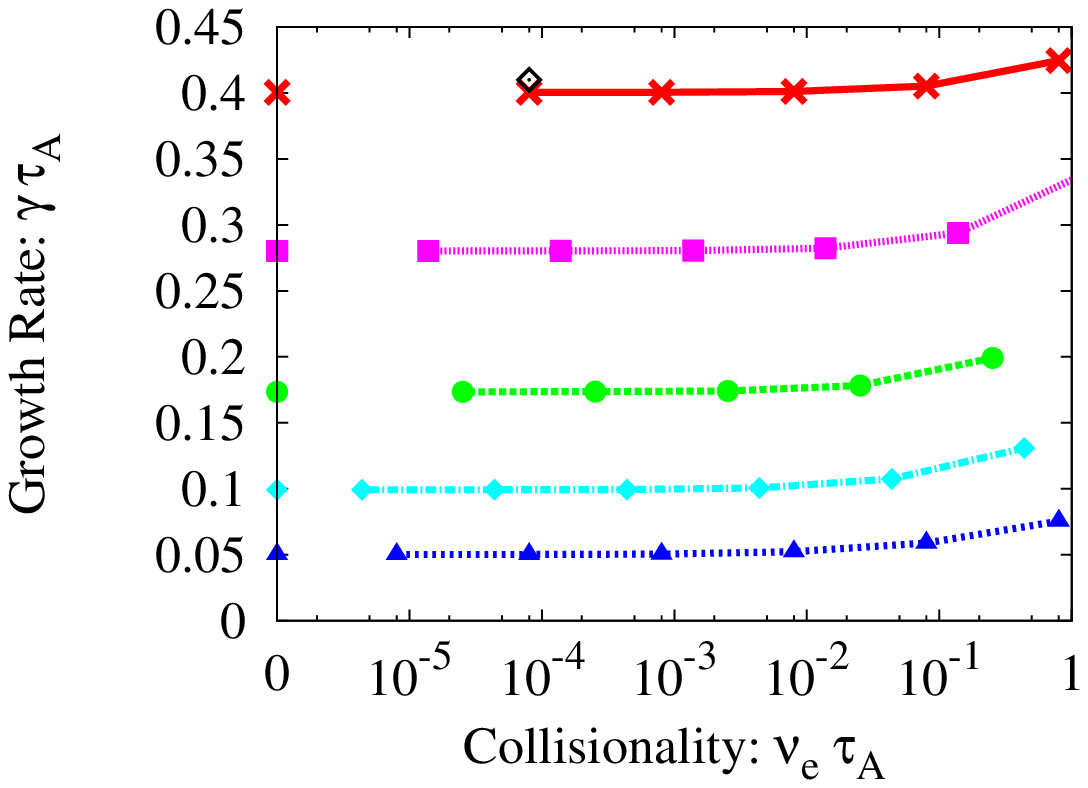}
  \includegraphics[scale=0.45]{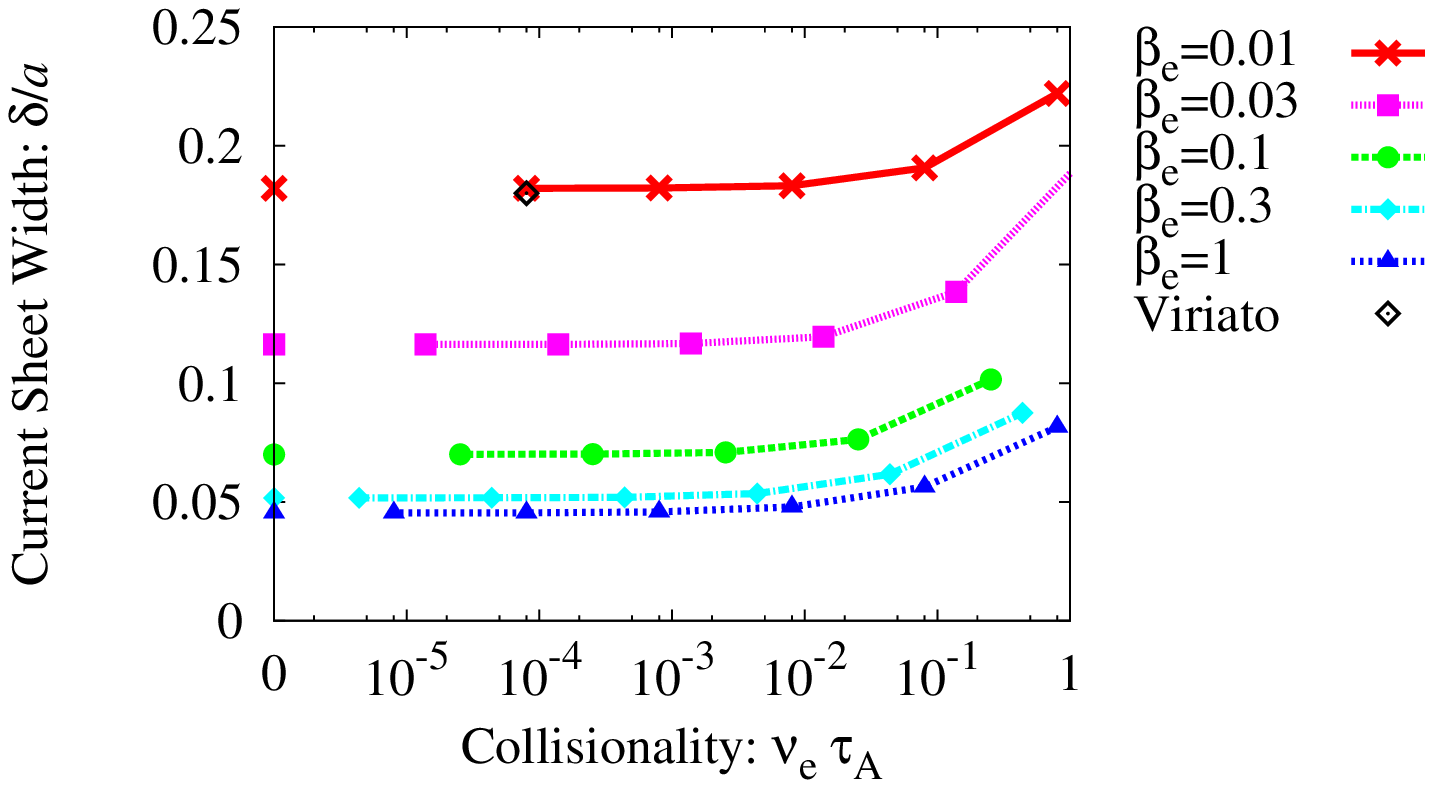}
  \vspace*{3em}
  \caption{\label{fig:linear}Collisionality scan of the linear growth
  rate and current layer width for different $\betae$
  values. In all cases we set $\nue=\nuee=\nuei$.
  The growth rate and current layer width obtained from
  the {\tt Viriato} code~\citep{LoureiroSchekochihinZocco_13} are also shown
  by black diamonds. The {\tt Viriato} values agree with the
  $\betae=0.01$ case.}
 \end{center}
\end{figure}

Shown in Fig.~\ref{fig:linear} are the growth rates ($\gamma$) and
current layer widths ($\delta$) as functions of collisionality.
For comparison, we also show the growth rate and current layer width for
$\nue \tauA = 8.0\times10^{-5}$ obtained from the reduced kinetic model
valid for $\betae\sim \me/\mi$ using the {\tt Viriato}
code~\citep{LoureiroSchekochihinZocco_13};
the agreement with the fully-gyrokinetic results at $\betae=0.01$ is
very good.

As collisionality decreases, the growth rate also decreases, and the
current layer becomes narrower. However, when collisionality becomes
sufficiently small, the growth rate and the current layer width
asymptote to constant values. This is the so-called
collisionless regime, even though the collision frequency is finite.
Note that, in the linear regime, small but finite
collisionality gives the same results as the exactly collisionless case
$\nue=0$.

\begin{figure}
 \begin{center}
  \includegraphics[scale=0.45]{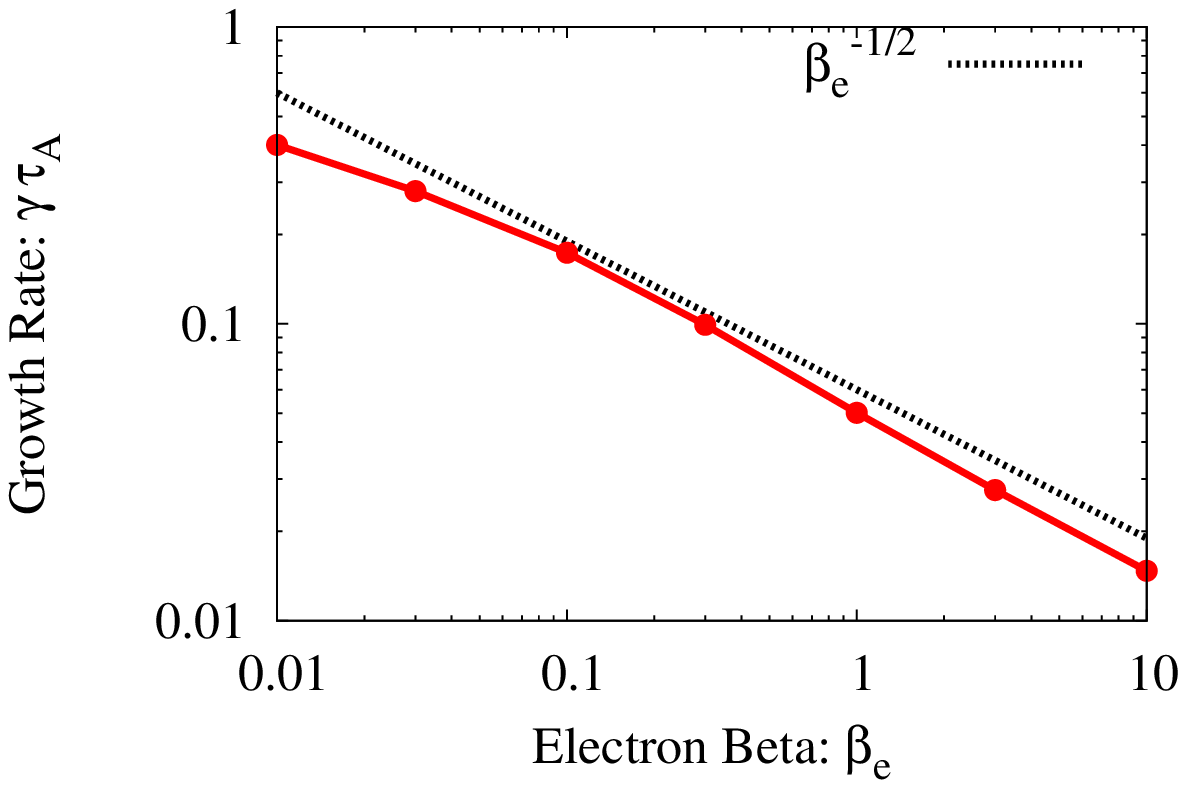}
  \includegraphics[scale=0.45]{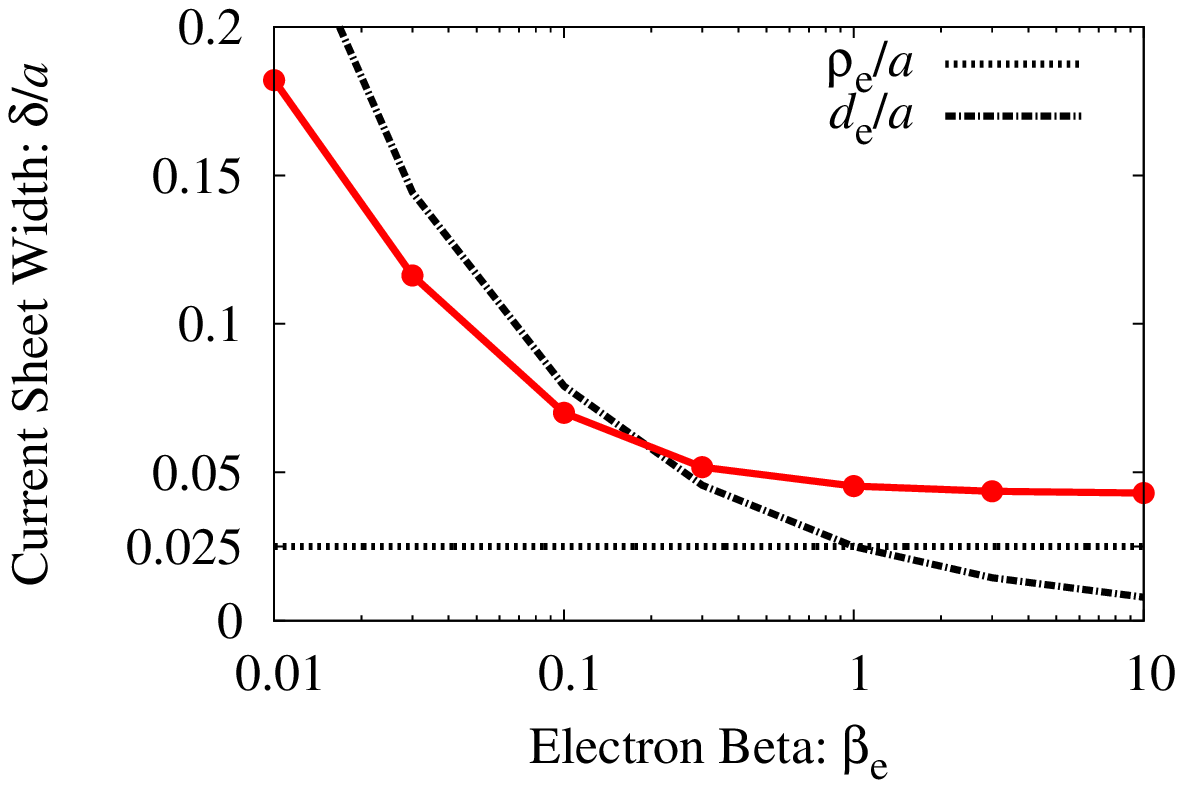}
  \vspace*{3em}
  \caption{\label{fig:lin_beta}$\betae$ dependence of the growth rate
  and the current layer width for the collisionless cases of
  Fig.~\ref{fig:linear}.}
 \end{center}
\end{figure}

Figure~\ref{fig:lin_beta} shows the growth rate and current layer width
for the collisionless cases against $\betae$. The growth rate decreases
with increasing $\betae$, scaling roughly as
$\betae^{-1/2}$ ($\propto \de$), as predicted by
\citet{Fitzpatrick_10} based on Braginskii's two-fluid model
\footnote{If $\betae$ is much smaller, or $\Delta'$ is much larger, the other
collisionless regime originally derived by \citet{Porcelli_91}, 
$\gamma\tauA\propto\betae^{-1/6}$ ($\gamma\tauA\propto\de^{1/3}$),
appears -- see \citet{NumataDorlandHowes_11}.}.
The
$\betae$ dependence of the current layer width indicates the physical
mechanism responsible for breaking the frozen-flux constraint. For
$\betae\ll 1$, the
electron Larmor radius is negligibly small compared with $\delta$, and
reconnection is mediated by electron inertia. As $\betae$ increases,
the current layer width decreases, and $\rhoe$ becomes comparable to
$\delta$ and $\de$. For $\betae>1$, $\rhoe$ becomes larger than $\de$,
and, thus, electron FLR effects overtake electron inertia as
the mechanism for breaking the flux-freezing condition.

\section{Nonlinear simulation results}
\label{sec:results}

We now report the results of nonlinear simulations in the collisionless
regime at different values of $\betae$.
We set $\nuee=\nuei=\nuii=0.8 \times
10^{-4} \sim 8 \times 10^{-4}$ (see Table~\ref{tbl:param})
\footnote[1]{
The ion-ion collisions do not significantly affect the reconnection
dynamics, and are just included to regularize velocity space structures
of the ion distribution function. An ion collision frequency equal to
the electron one is assumed to be sufficient since the ion phase mixing
is at most as strong as that of electrons.
Ideally, a similar convergence test for ions as electrons shown in
Appendix~\ref{sec:vconv} should be performed.
}.

In all simulations reported here, the number of grid points in the $x$,
$y$ directions are $N_x=256$, $N_y=128$ (subject to the 2/3's rule for
de-aliasing), except for $\betae=1$, where $N_x=512$ such that the linear
stage can be adequately resolved. In velocity space, we use
$N_{\lambda}=N_{E}=64$ for all cases, as required by the convergence
test described in Appendix~\ref{sec:vconv}.

\subsection{Temporal evolution of magnetic reconnection}
\label{sec:tevo}

\begin{figure}
 \begin{center}
  \includegraphics[scale=0.6]{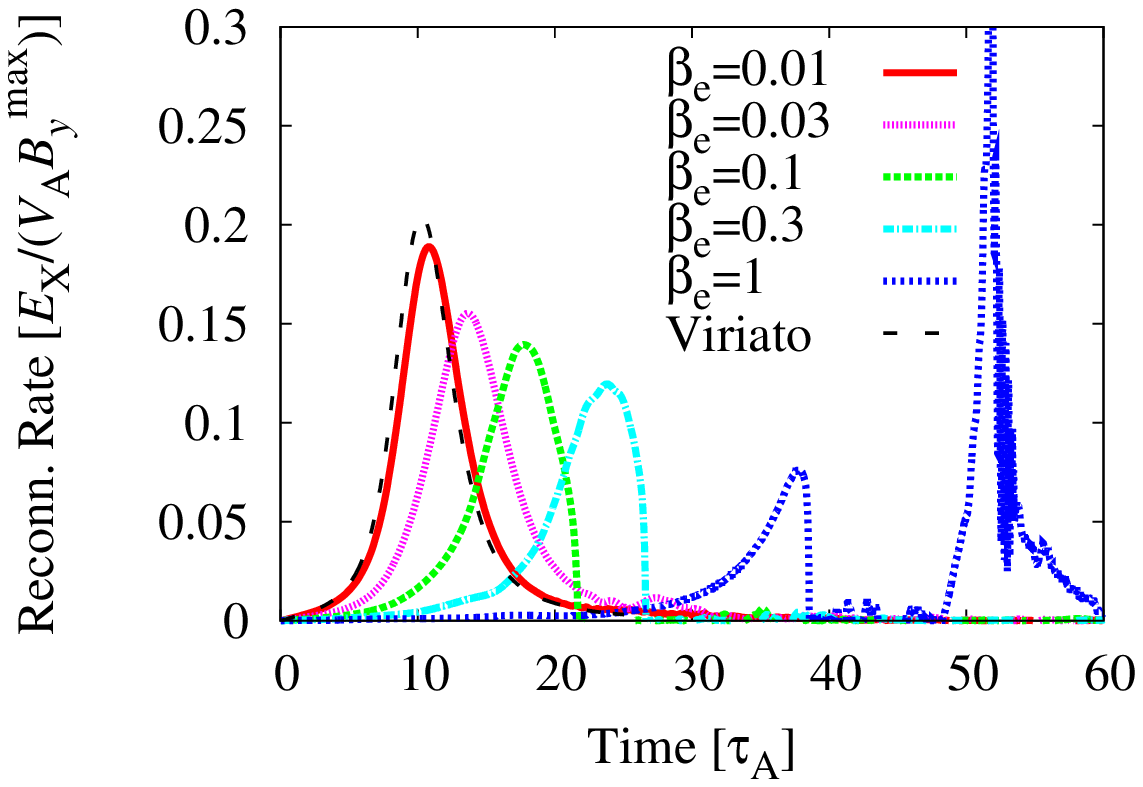}
  \includegraphics[scale=0.4]{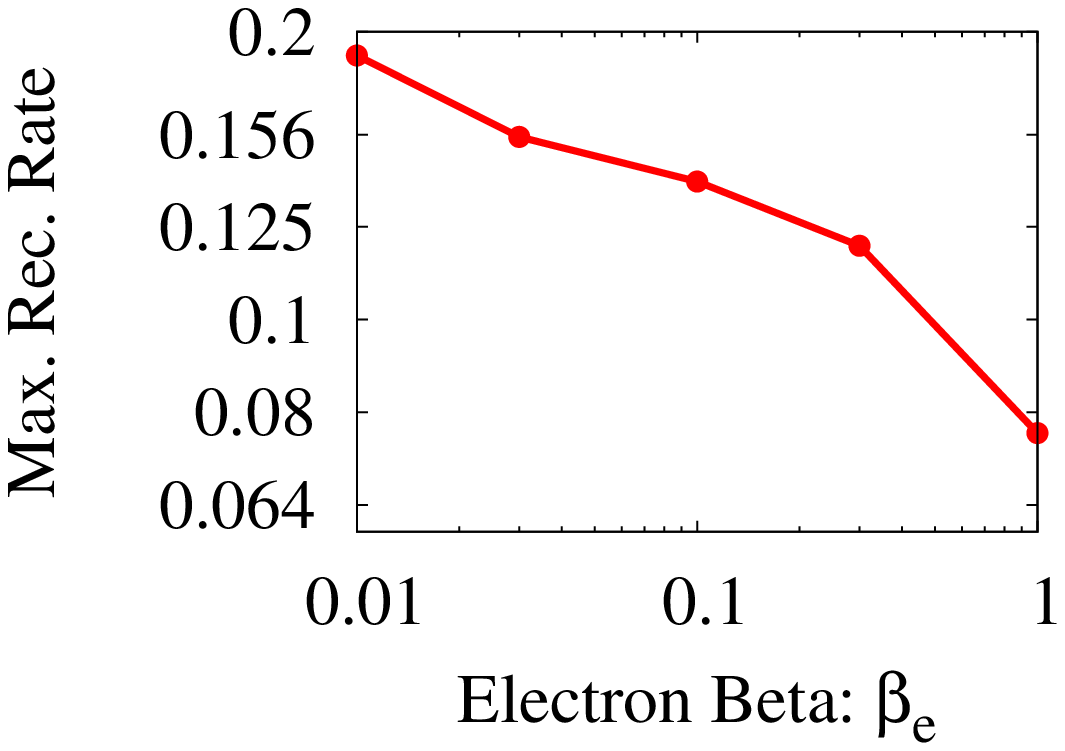}
  \vspace*{3em}
  \caption{\label{fig:recrate}Left panel: The electric field at the
  center of the domain as a function of time.
  At early times, the center of the domain is an $X$ point and $E_{X}$
  represents the reconnection rate. The reversal of the electric field
  observed for $\betae\gtrsim0.1$ indicates the conversion of the
  $X$ point into an $O$ point, \ie the current sheet is unstable to
  plasmoid formation. From that moment onwards, $E_{X}$ ceases to
  represent the reconnection rate.
  \revise{The second peak for $\betae=1$ indicates that the plasmoid
  moves away from the center of the domain because of the numerical
  noise.}
  Right panel: Maximum reconnection rate as a function of $\betae$.
  \revise{(In the $\betae=1$ case, we plot the maximum reconnection rate
  achieved prior to plasmoid formation.)}
  }
 \end{center}
\end{figure}

Figure~\ref{fig:recrate} shows the time evolution of the electric field
at the $X$ point $(x,y)=(L_x/2,L_y/2)$ ($E_{X}$) as a measure of the
reconnection rate for different values of $\betae$.
The peak reconnection rate values that we find ($\gtrsim 0.1$) are
comparable with the standard value usually reported in the literature in
the no-guide-field case; in addition, we find that the peak reconnection
rate is a weakly decreasing function of $\betae$.
The reconnection rate from {\tt Viriato} is $\sim0.2$ again showing
very good agreement with the $\betae=0.01$ case.
For higher $\betae$ cases, the electric field drops sharply
shortly after the peak reconnection rate is achieved, eventually
reversing sign; this occurs because the current sheet becomes
unstable to the secondary plasmoid
instability~\citep{LoureiroSchekochihinCowley_07,LoureiroSchekochihinUzdensky_13},
with the $X$ point 
becoming an $O$ point instead -- see Fig.~\ref{fig:diss_b1_e} for the
$\betae=1$ case.
In the simulations shown in this work, the
secondary island stays at the center of the domain because there is no
asymmetry in the direction of the reconnecting field lines (\ie the
outflow direction).
\revise{For the $\betae=1$ case, however, it eventually moves toward the 
primary island due to the numerical noise as indicated by the
sharp spike of $E_{X}$ around $t/\tauA=52$.}

\begin{figure}
 \begin{center}
  \includegraphics[scale=0.5]{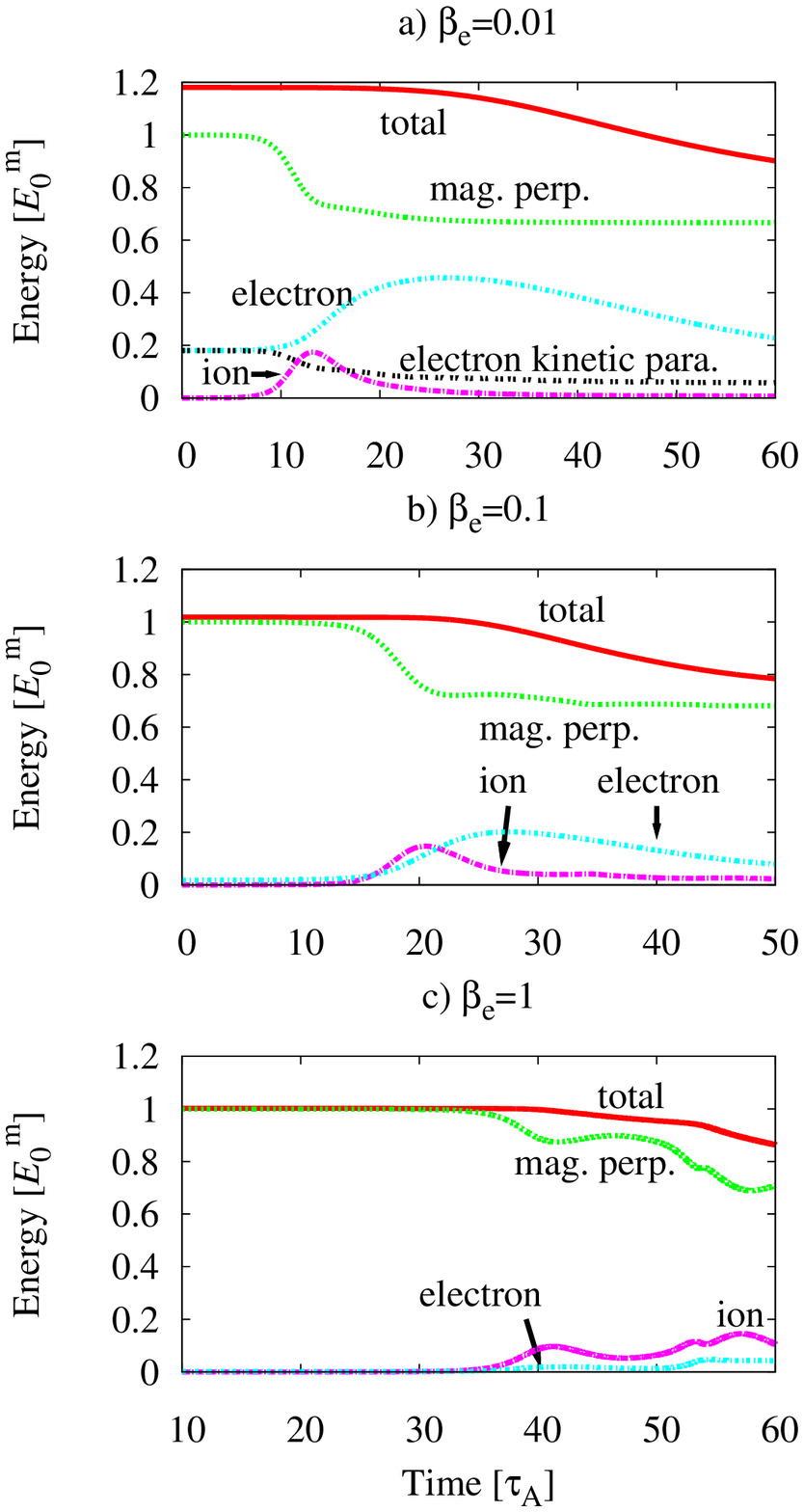}
  \includegraphics[scale=0.5]{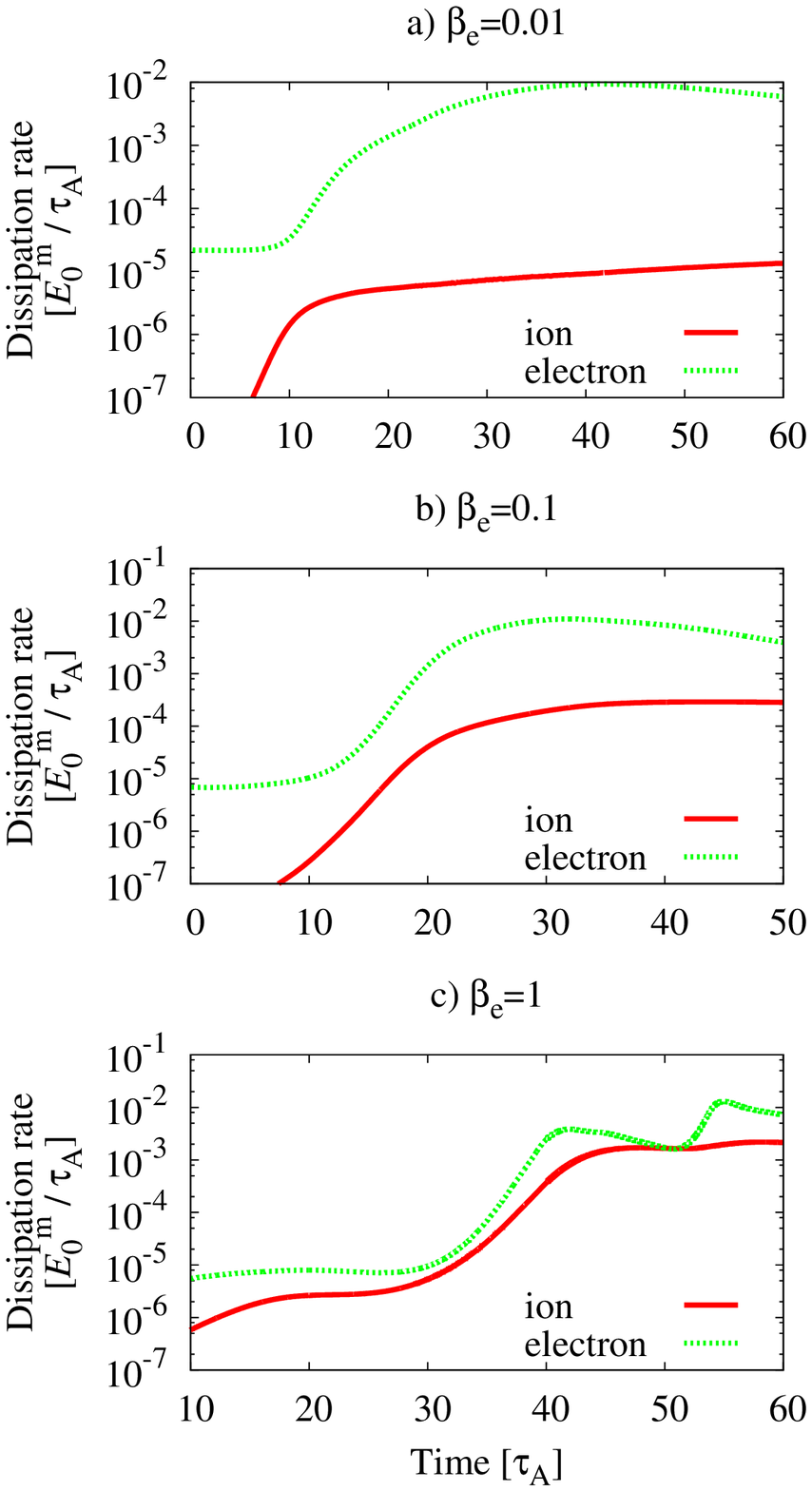}
  \caption{\label{fig:energy}Time evolution of the energy components
  (left) and the dissipation rate of ions and electrons (right) for
  $\betae=0.01, 0.1, 1$. Note the time for the $\betae=1$ case is shifted
  by $t/\tauA=10$. The energies are normalized by the initial magnetic
  field energy.}
 \end{center}
\end{figure}

To detail the energy conversion processes taking place in our simulations
we plot in Fig.~\ref{fig:energy} the time evolution of each component of
the generalized energy \eqref{eq:gen_energy} normalized by the initial
magnetic energy ($E_{0}^{\mathrm{m}}$),
\revise{except the parallel magnetic energy
($E_{\parallel}^{\mathrm{m}}$) as it is almost zero for all cases.}
During magnetic reconnection, the magnetic energy is converted to the
 particle's energy reversibly. Looking at the $\betae=0.01$ case we see
 that, first, the ion $\EB$ drift flow is excited, thereby increasing
 the ion energy.
 \revise{Then, electrons exchange energies with the excited fields. The
 parallel electric field accelerates or decelerates electrons depending
 on their orbit and energy, but the net work done on electrons for this
 case is positive. Even}
 though electrons lose parallel kinetic energy by resistive
 dissipation,
 they gain energy via the phase-mixing process \nuno{as will be
 explained in detail shortly}, and store the increased
 energy in the form of temperature fluctuations and higher order 
 moments. Collisionally dissipated energy (decrease of the total energy)
 is about 1\% of the initial magnetic energy after dynamical process has
 almost ended ($t/\tauA=25$).

Also shown in Fig.~\ref{fig:energy}, right panel, are the dissipation rates of
electrons and ions as determined from \eqref{eq:dissipation_rate}.
The dissipation rate is normalized by the initial
magnetic energy divided by the Alfv\'en time,
$E^{\mathrm{m}}_{0}/\tauA$. For $\betae=0.01$, the electron energy
dissipation starts to grow rapidly when the maximum reconnection rate is 
achieved. It stays large long after the dynamical stage, and an
appreciable amount of the energy is lost at late times. The ion
dissipation is negligibly small compared with the electron's for this
case.


With regards to energy partition, the most important effect to notice as
$\betae$ increases is the decrease in the electron energy gain. For 
the $\betae=1$ case, the released magnetic energy is contained mostly by
ions \nuno{as shown in Fig.~\ref{fig:energy}, bottom left}.

\revise{The plasmoid formation and ejection complicate the evolution of
energies and dissipations for the $\betae=1$ case. \nuno{From the
bottom-left panel of Fig.~\ref{fig:energy},} we notice that there is a
slight increase of the magnetic energy during the plasmoid growth phase
($40\lesssim t/\tauA \lesssim 50$). The electron dissipation decreases
in this phase, and is re-activated when the plasmoid is ejected
(Fig.~\ref{fig:energy}, bottom right).
\nuno{This secondary heating of electrons is
associated with the newly formed $X$ point as will be shown in
Fig.~\ref{fig:diss_b1_plasmoid}.}}

\begin{figure}
 \begin{center}
  \includegraphics[scale=0.6]{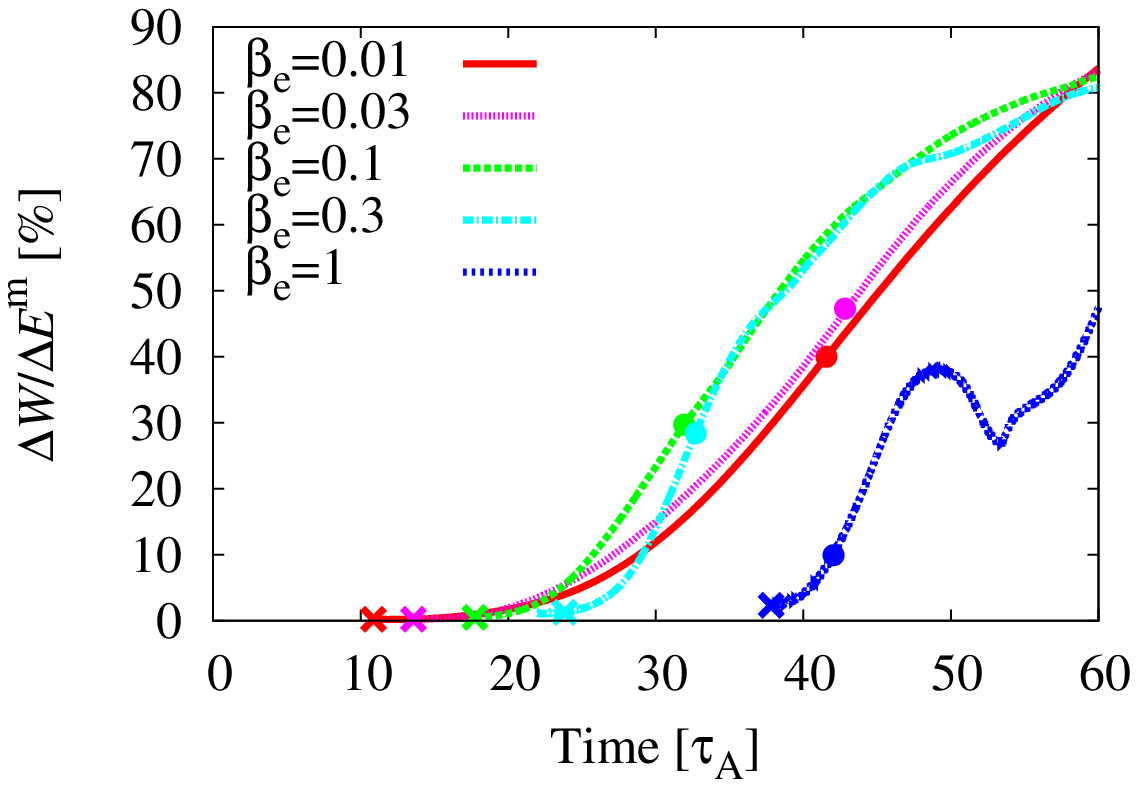}
  \vspace*{2em}
  \caption{\label{fig:fraction}Ratio of the dissipated energy to the
  released magnetic energy as a function of time.
  The cross ($\times$) and dot ($\bullet$) symbols on each line denote the
  times of the maximum reconnection rate and the maximum electron
  dissipation rate, respectively.
  \revise{The plots are shown only in the nonlinear regime since both the
  dissipation and magnetic energy release are almost zero in the
  linear regime.}
}
 \end{center}
\end{figure}

A significant fraction of the released magnetic energy is collisionally
dissipated. The fraction of the dissipated energy
[$\Delta W=W(t)-W(0)$] to the released magnetic energy [$\Delta
E^{\mathrm{m}}=E^{\mathrm{m}}(t)-E^{\mathrm{m}}(0)$] is shown in
Fig.~\ref{fig:fraction}. The cross ($\times$) and dot ($\bullet$) symbols
on each line indicate the times of the maximum reconnection rate 
and the maximum electron dissipation rate, respectively.
\revise{We show $\Delta W/\Delta E^{\mathrm{m}}$ only in the nonlinear
regime since both $\Delta W$ and $\Delta E^{\mathrm{m}}$ are almost zero
in the linear regime (see Fig.~\ref{fig:energy}).}
The total energy dissipation is determined by the local strength of
phase mixing (the magnitude of the integrand of \eqref{eq:dissipation_rate}
integrated over only the velocity coordinates at each position), the
total area where phase mixing is active, and the duration of phase
mixing. Therefore, 
the energy conversion from the magnetic to thermal energy (\ie
how much of the released magnetic energy is dissipated from the system
($\Delta W/\Delta E^{\mathrm{m}}$)) as a function of time has no clear
dependence on $\betae$.
\revise{For the $\betae=1$ case, in particular, $\Delta W/\Delta
E^{\mathrm{m}}$ evolves in a complex manner because the evolution of
energies and dissipations are significantly altered by the plasmoid as
described above.}

\begin{figure}
 \begin{center}
  \includegraphics[scale=0.6]{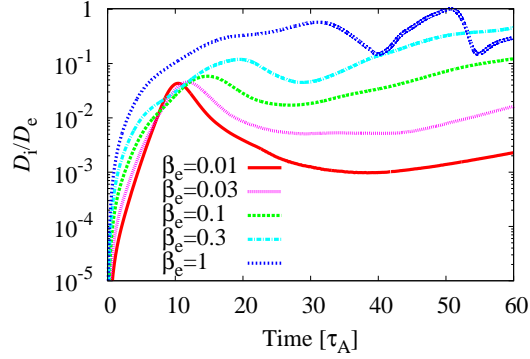}
  \vspace*{2em}
  \caption{\label{fig:heating_ratio}Ratio of the dissipation rate of
  ions to electron for five different $\betae$ values. Ion heating is
  comparable to electron heating for $\betae\sim1$.}
 \end{center}
\end{figure}

As $\betae$ is increased, the ion dissipation also becomes large. As
shown in Fig.~\ref{fig:heating_ratio}, the ratio of the energy
dissipation of ions to electrons, $D_{\mathrm{i}}/D_{\mathrm{e}}$,
becomes approximately unity for the $\betae=1$ case implying that ion
heating is as significant as electron heating when $\betae \sim 1$.

\subsection{Phase mixing and energy dissipation}
\label{sec:phasemixing}

\begin{figure}
   \begin{center}
    \includegraphics[scale=0.22]{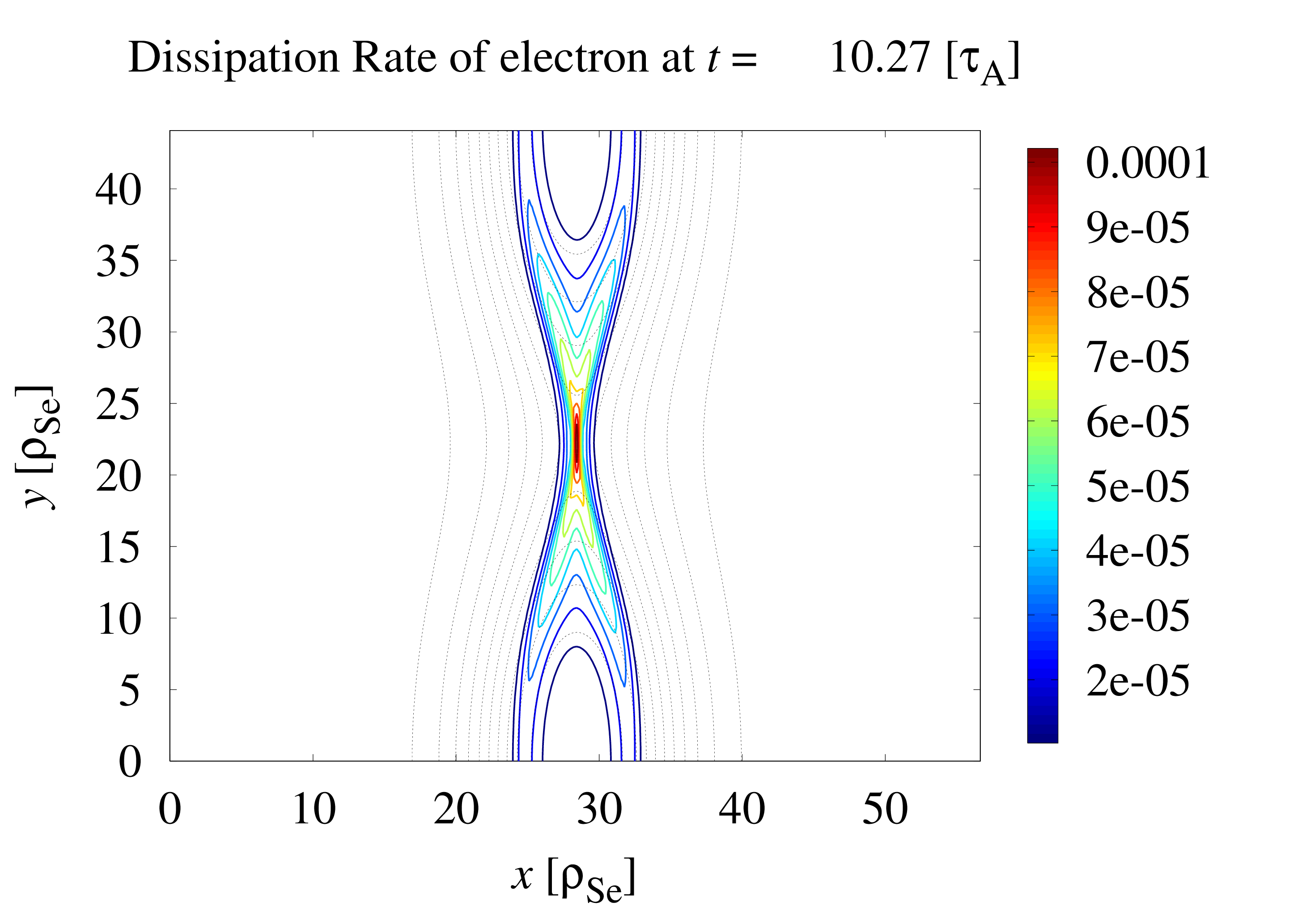}
    \includegraphics[scale=0.22]{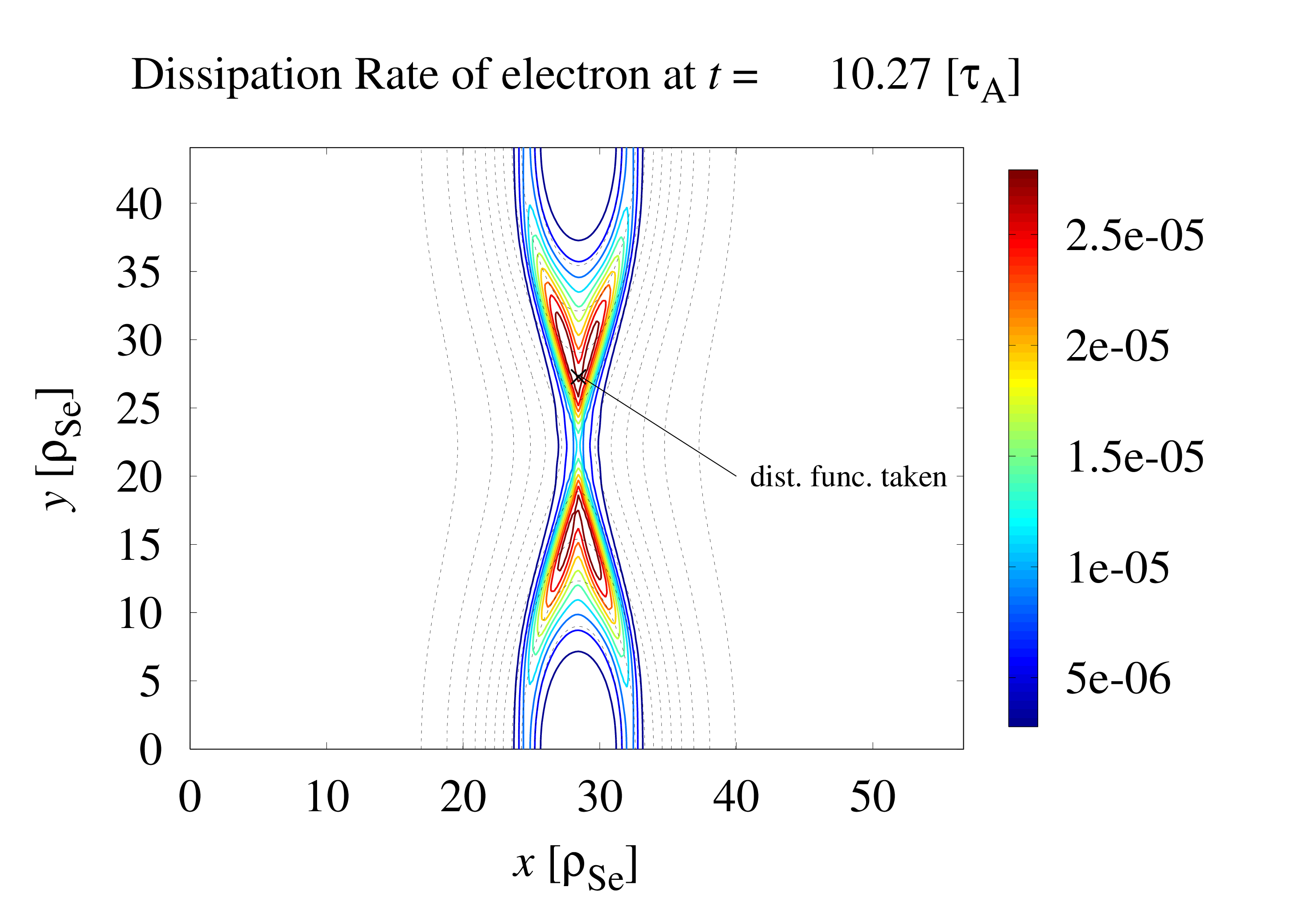}
    \includegraphics[scale=0.22]{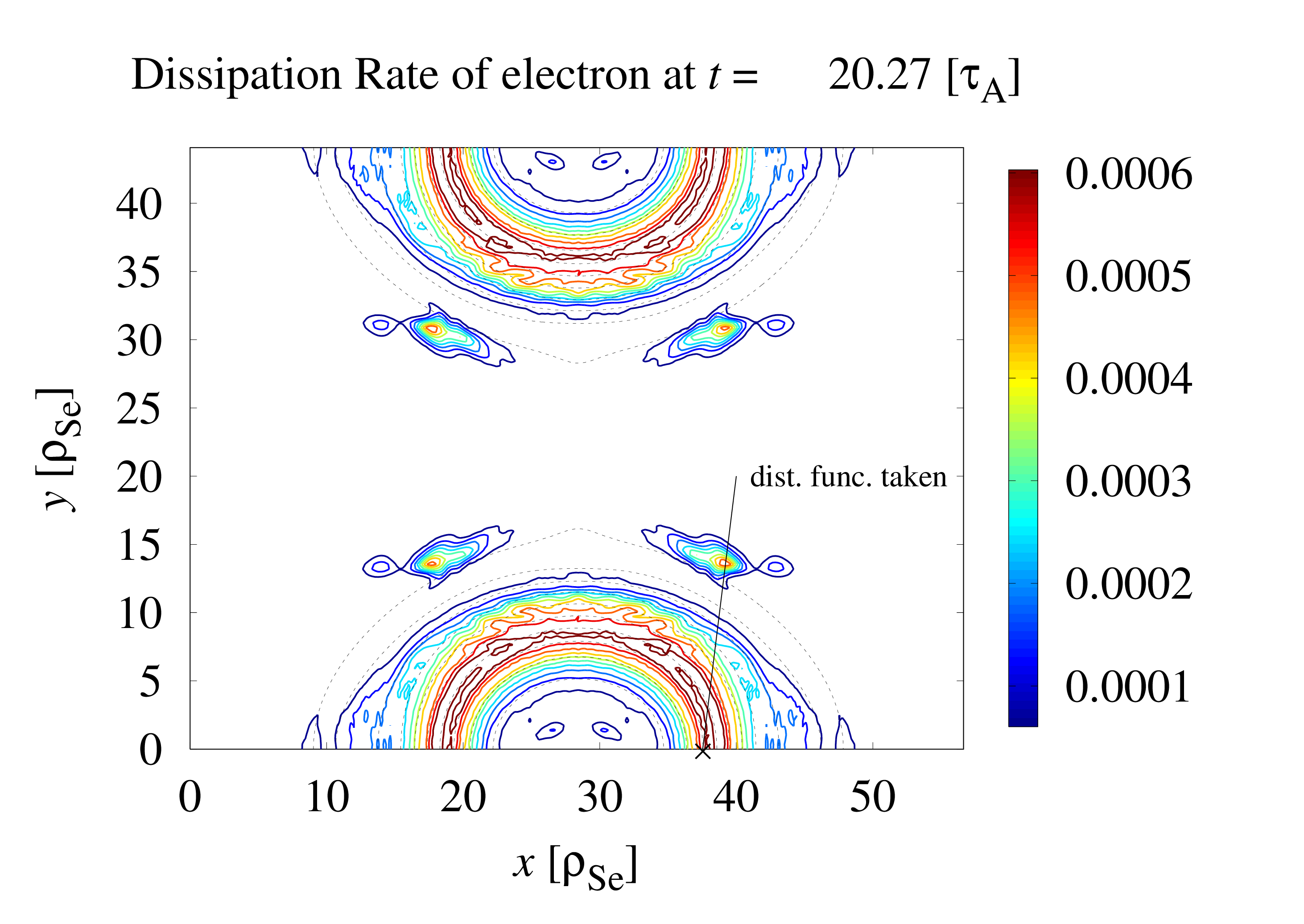}
    \caption{\label{fig:diss_b0.01}Spatial distributions of the
    dissipation rate of electrons
    ($D_{\mathrm{e}}/(E^{\mathrm{m}}_0/\tau_A)$) for $\betae=0.01$. 
    The top two panels show the dissipation rate at $t/\tauA=10$ where
    the reconnection rate is maximum. The left panel includes the full
    dissipation, while the right does not include the resistivity
    component. The bottom figure shows the dissipation rate at
    $t/\tauA=20$.}
   \end{center}
  \end{figure}

Figure~\ref{fig:diss_b0.01} shows spatial distributions of the
dissipation rate of electrons for $\betae=0.01$.
The top two panels show the dissipation rate of the electron energy at
$t/\tauA=10$, \ie when the reconnection rate is 
maximum. At this time, the electron dissipation is localized near the
reconnection point. If we subtract the first two velocity moments from
the distribution function, which correspond to perturbations of the
density and bulk flow in the $z$ direction
\revise{(\ie the dissipation rate based on $h_s'$ defined in
\eqref{eq:hermite})},
we find that the dissipation occurs just downstream of the reconnection
site. This indicates that the dissipation at the $X$ point is the
resistivity component, corresponding to the decay of the initial
electron current. The contribution of higher order
moments to the dissipation spreads along the separatrix.
The energy lost via Ohmic heating at this stage is much smaller than the
dissipation due to 
phase mixing that occurs later, and is not seen in later times.
(In all other plots of the spatial distribution of $D_{\mathrm{e}}$,
the resistivity component is negligible.)

After the dynamical reconnection phase ended, most of the available flux
has been reconnected and a large island is formed.
Electron dissipation is large inside the island
(bottom panel of Fig.~\ref{fig:diss_b0.01}), and continuously
increases in time -- see Fig.~\ref{fig:energy}, top right.

\begin{figure}
  \begin{center}
   \includegraphics[scale=0.3]{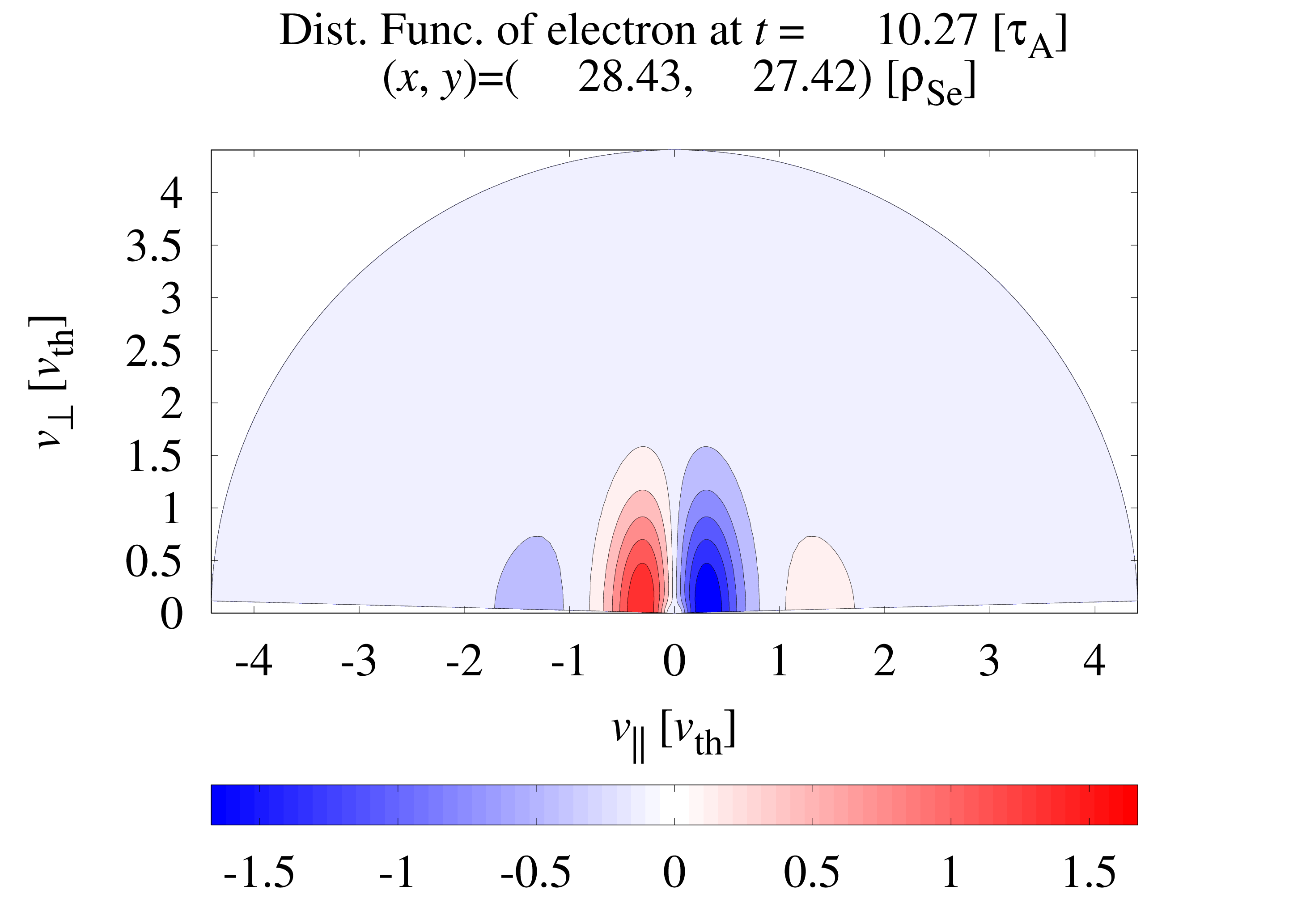}
   \includegraphics[scale=0.3]{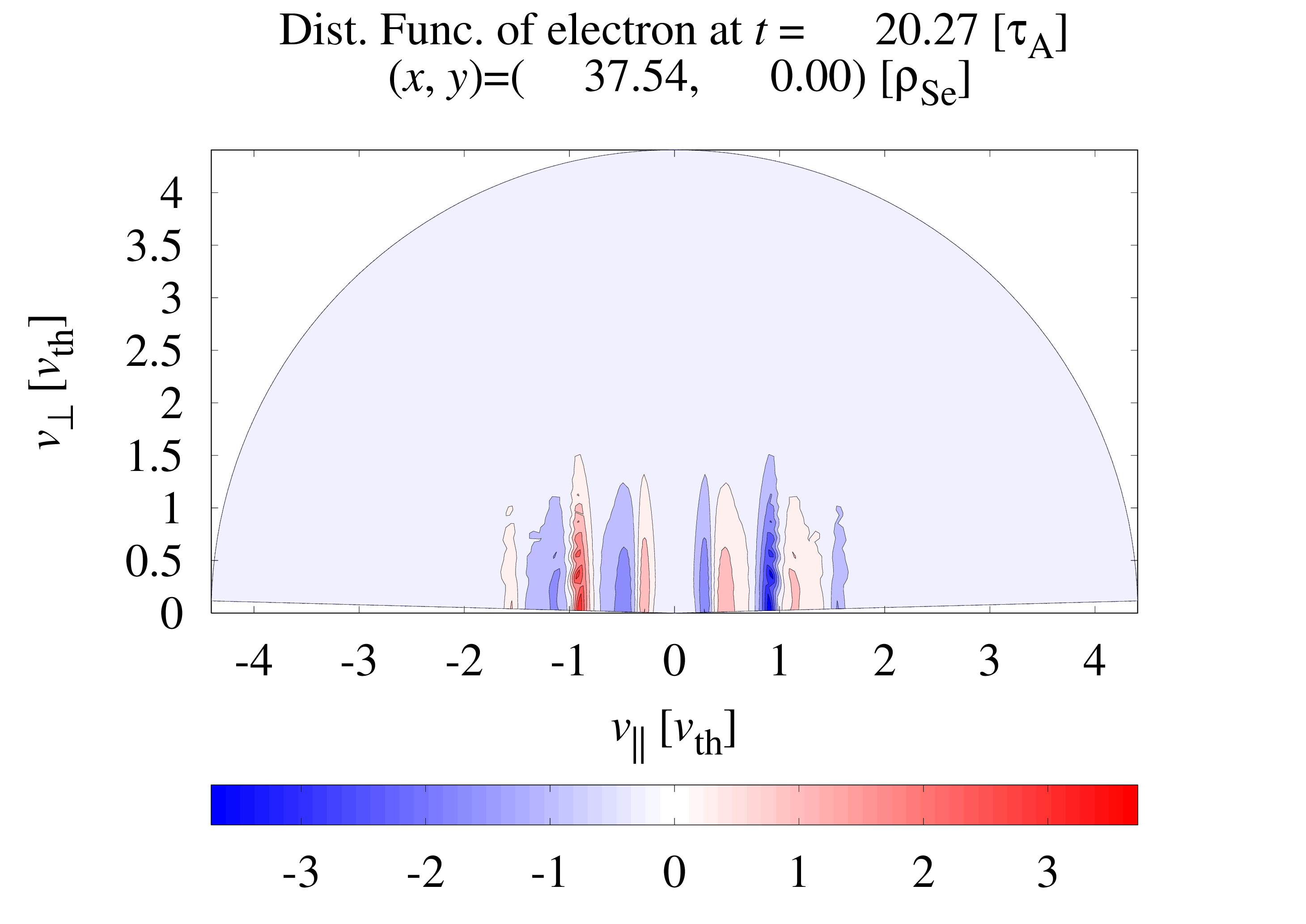}
   \caption{\label{fig:dstfne_b0.01}Velocity space structures of electron
   distribution function, without the first two velocity moments, for
   $\betae=0.01$, normalized by 
   $\varepsilon n_{0}/(\sqrt{\pi}\vthe)^{3}$. Distribution functions
   are taken at the strongest dissipation point.}
  \end{center} 
 \end{figure}

In the regions where the dissipation rate is large, we expect the
distribution function to show oscillatory structures due to phase mixing
(the local strength of dissipation is roughly proportional to $\nu
h^{2} \delta 
v^{-2}$ where $\delta v$ denotes the scale length of the distribution
function in velocity space). Figure~\ref{fig:dstfne_b0.01} shows electron
distribution functions, without the first two velocity moments, taken
where the non-resistive part of dissipation
rate is largest for $t/\tauA=10, 20$. The distribution function is
normalized by $\varepsilon n_{0}/(\sqrt{\pi} \vthe)^{3}$.
The distribution function
only has gradients in the $v_{\parallel}$ direction, indicating that
parallel phase mixing is occurring, and the scale length $\delta v$
becomes smaller as time progresses. The phase-mixing structures develop
slowly compared with the time scale of the dynamical reconnection
process, therefore the energy dissipation peaks long after the peak
reconnection rate~\citep{LoureiroSchekochihinZocco_13}.

Phase mixing is significant when the Landau resonance condition
is met: Electrons moving along the reconnected field lines feel
variations of the electromagnetic fields evolving with the velocity
$\sim\VA$ in the perpendicular plane.
Since the magnetic field lies dominantly in the $z$ direction, but has
a small component in the $x-y$ plane, when electrons run along the field
lines with $v_{\parallel}$, electrons also move in the $x-y$ plane with
$v_{\parallel}\left(B_{\perp}/B_{z0}\right)$, thus the resonance
condition is given by $v_{\parallel} \left(B_{\perp}/B_{z0}\right) \sim\VA$.
For Fig.~\ref{fig:dstfne_b0.01}, phase mixing is most pronounced
around $v_{\parallel}/\vthe\sim1$, which agrees with the resonance
condition for $\betae=0.01$ and  the mass ratio
$\sigma=0.01$. Therefore, electrons with $v_{\parallel}/\vthe\sim1$ can
effectively exchange energies with the fields ($\phi$ in this case
because $\delta B_{\parallel}$ is small).

\revise{Parallel heating driven by the curvature drift through the Fermi
acceleration mechanism suggested by \citet{DrakeSwisdakChe_06} and
\citet{DahlinDrakeSwisdak_14} is negligible for the strong-guide-field
case discussed here simply because the curvature drift is small compared
with the thermal velocity.}

\begin{figure}
  \begin{center}
   \includegraphics[scale=0.22]{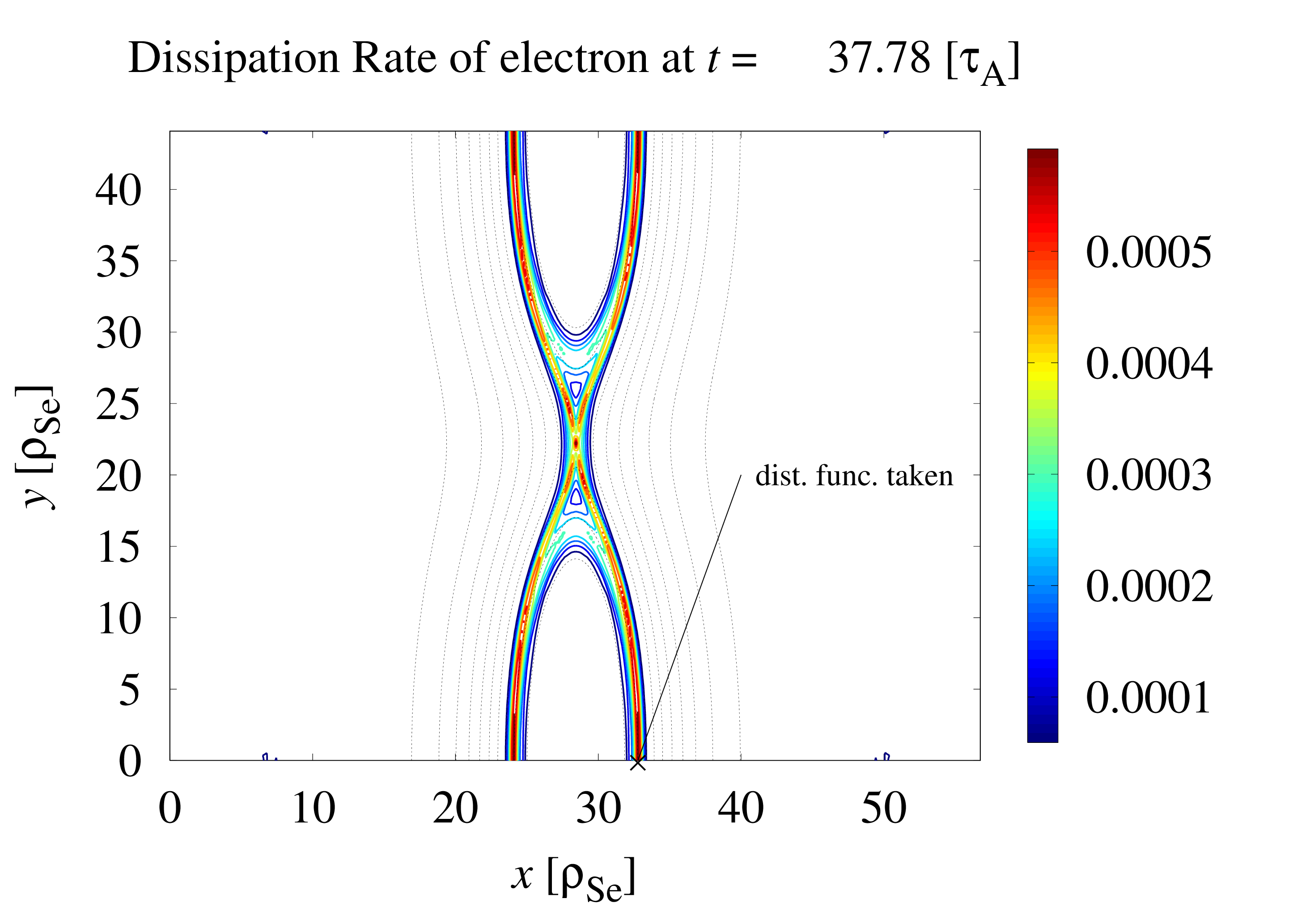}
   \includegraphics[scale=0.22]{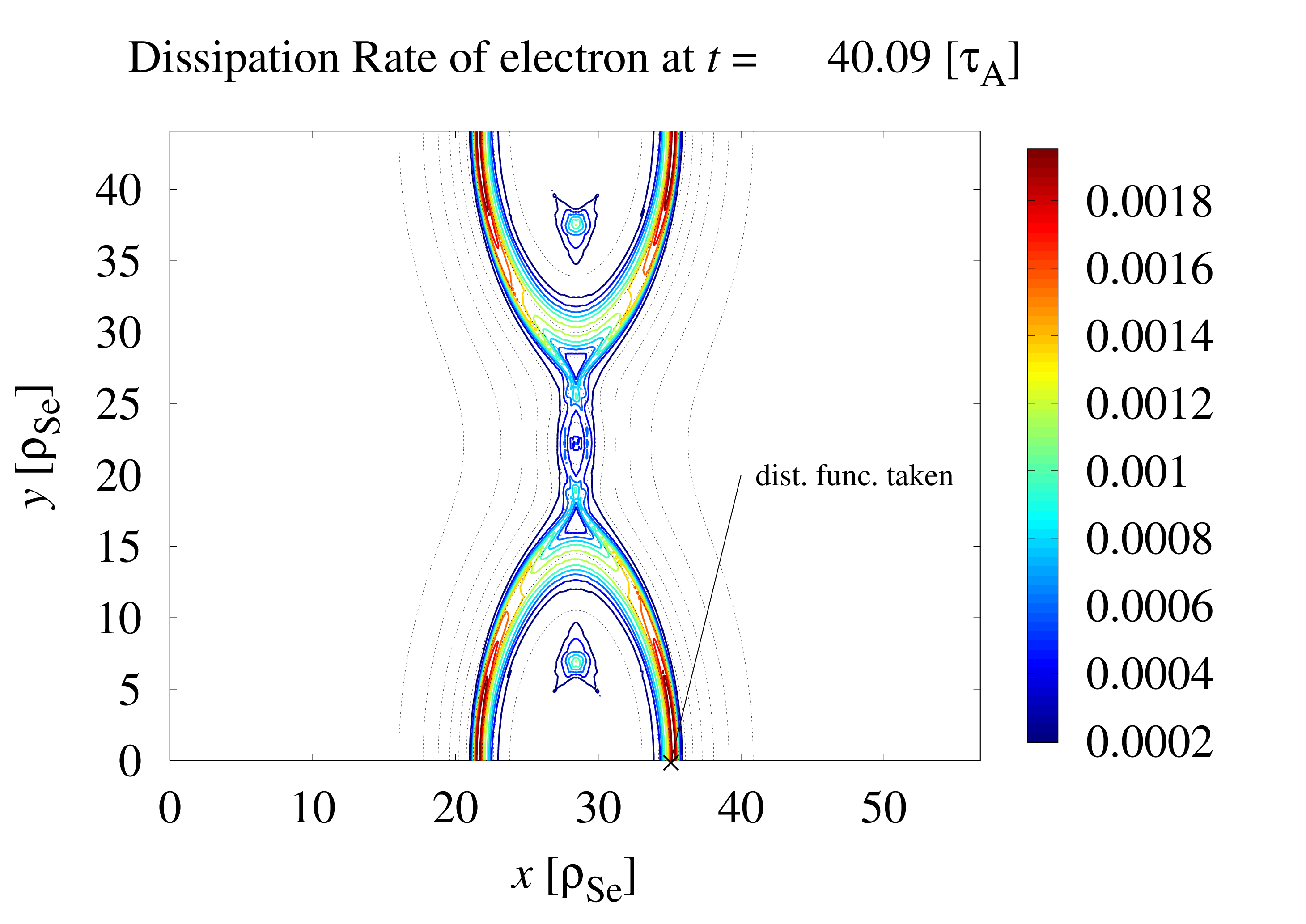}
   \caption{\label{fig:diss_b1_e}Spatial distribution of the dissipation
   rate of electrons ($D_{\mathrm{e}}/(E^{\mathrm{m}}_0/\tau_A)$) for $\betae=1$.}
  \end{center}
 \end{figure}

\begin{figure}
  \begin{center}
   \includegraphics[scale=0.3]{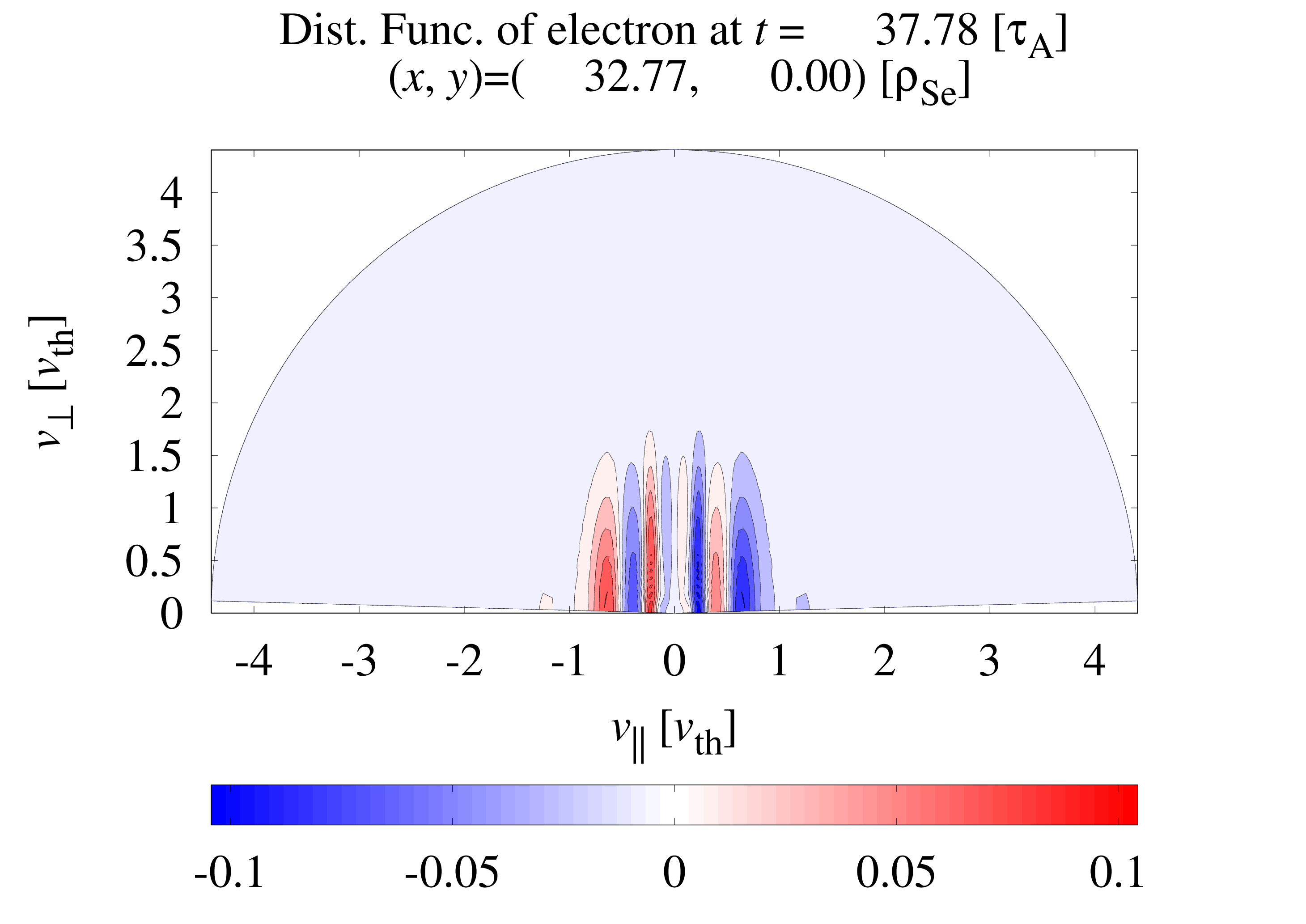}
   \includegraphics[scale=0.3]{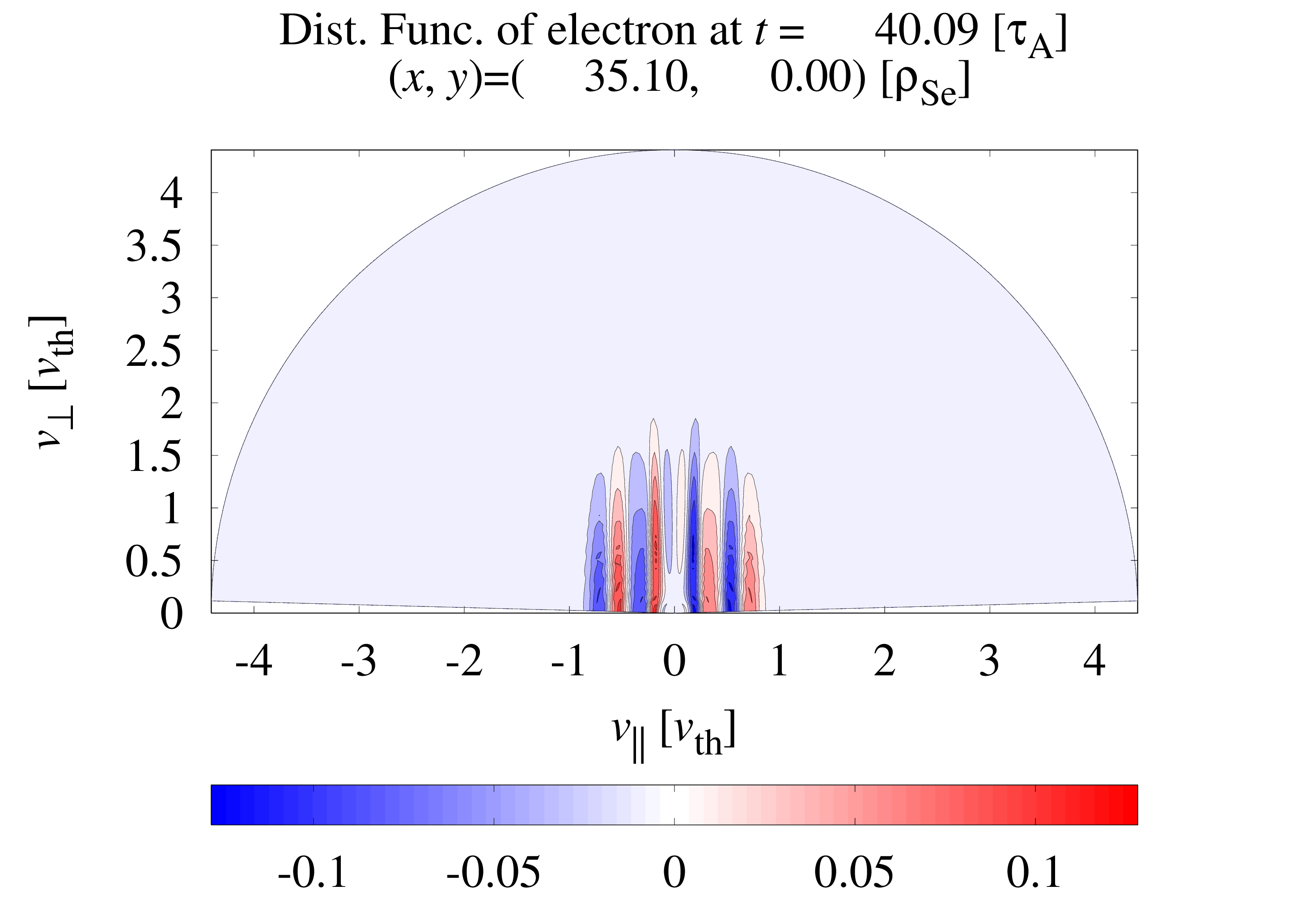}
   \caption{\label{fig:dstfn_b1_e}Velocity space structures of electron
   distribution function, without the first two velocity moments, for
   $\betae=1$, normalized by
   $\varepsilon n_{0}/(\sqrt{\pi}\vthe)^{3}$.}
  \end{center}
 \end{figure}

\begin{figure}
  \begin{center}
   \includegraphics[scale=0.22]{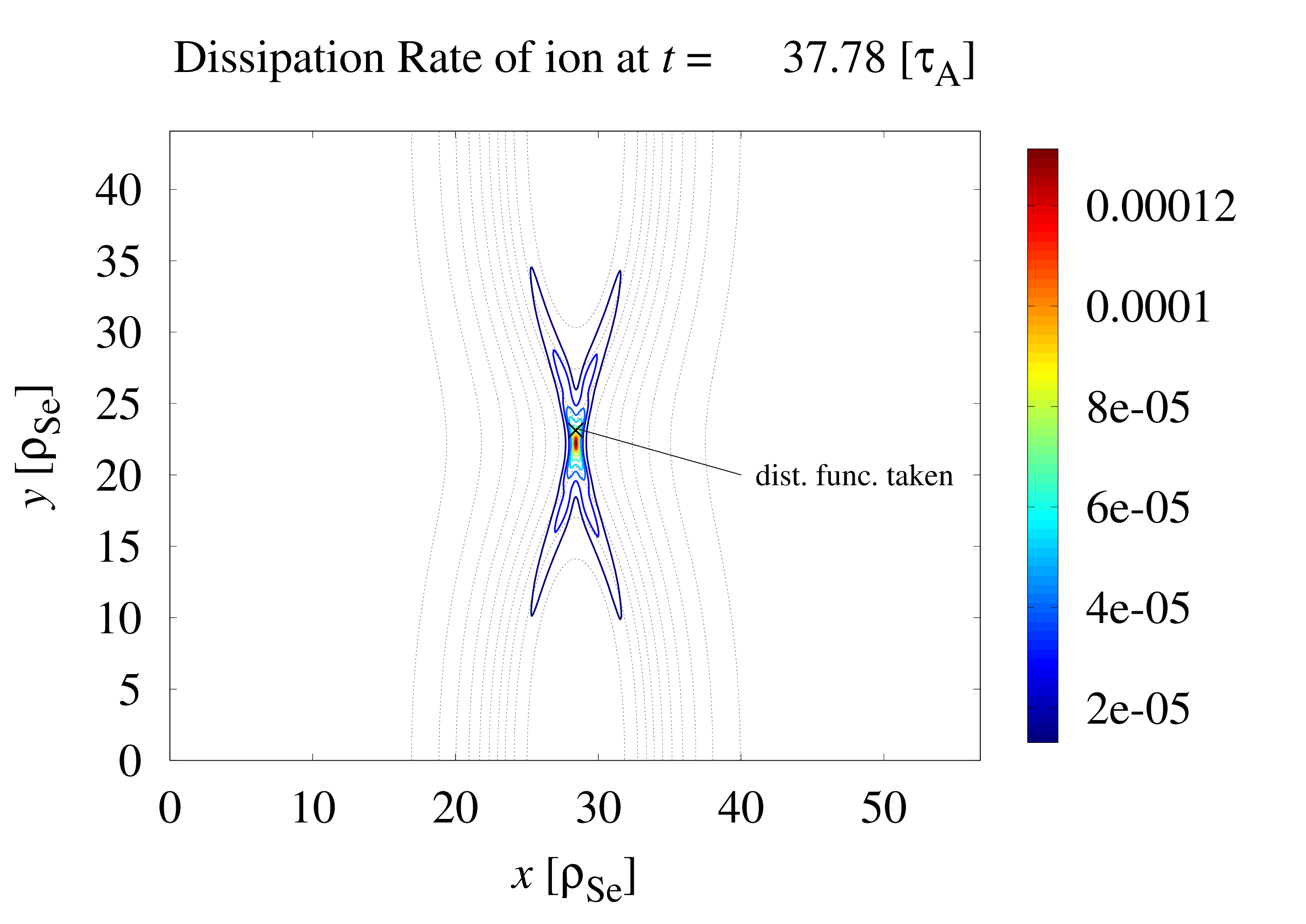}
   \includegraphics[scale=0.22]{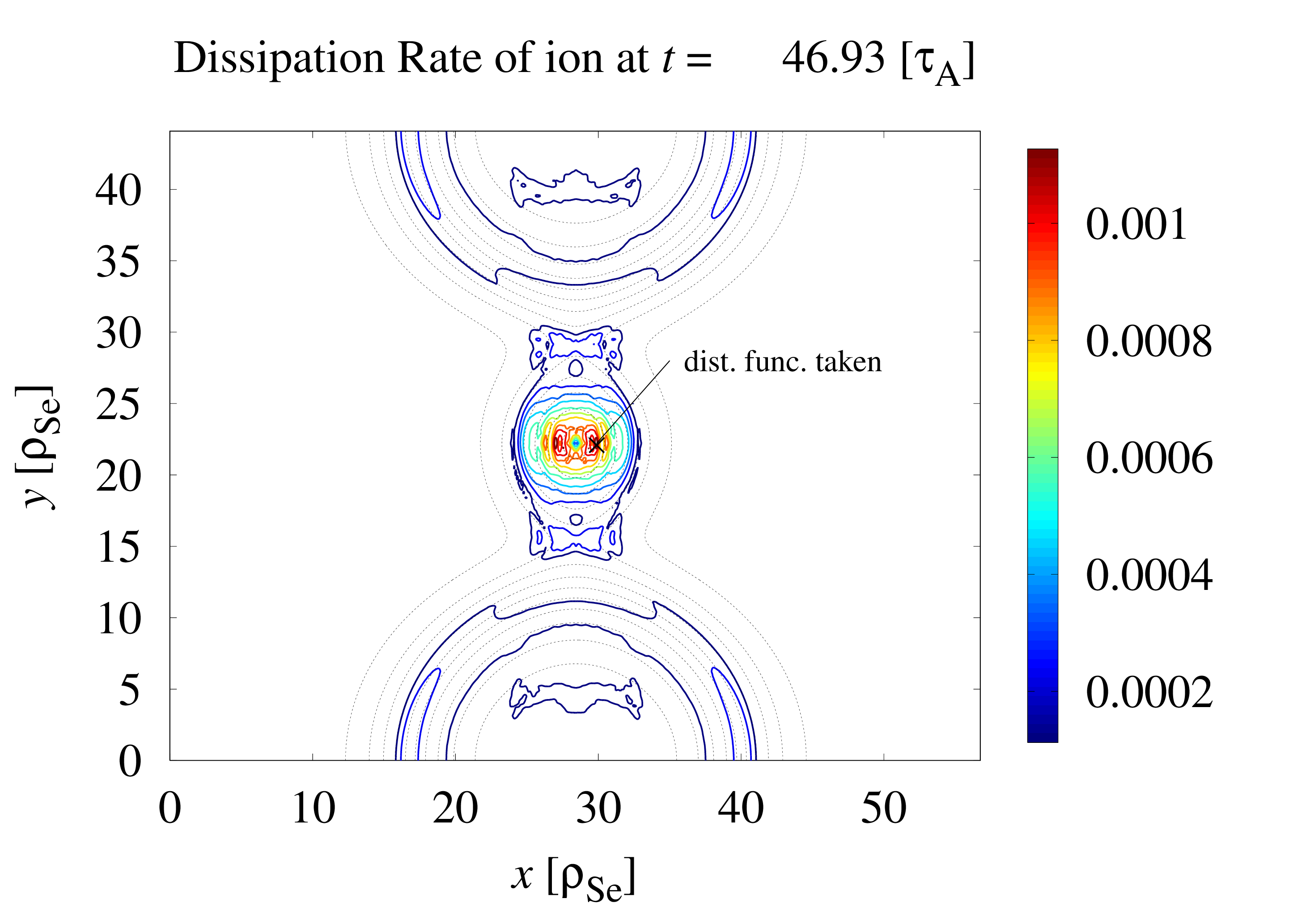}
   \caption{\label{fig:diss_b1_i}Spatial distribution of the dissipation
   rate of ions ($D_{\mathrm{i}}/(E^{\mathrm{m}}_0/\tau_A)$) for $\betae=1$.}
  \end{center}
 \end{figure}

\begin{figure}
  \begin{center}
   \includegraphics[scale=0.3]{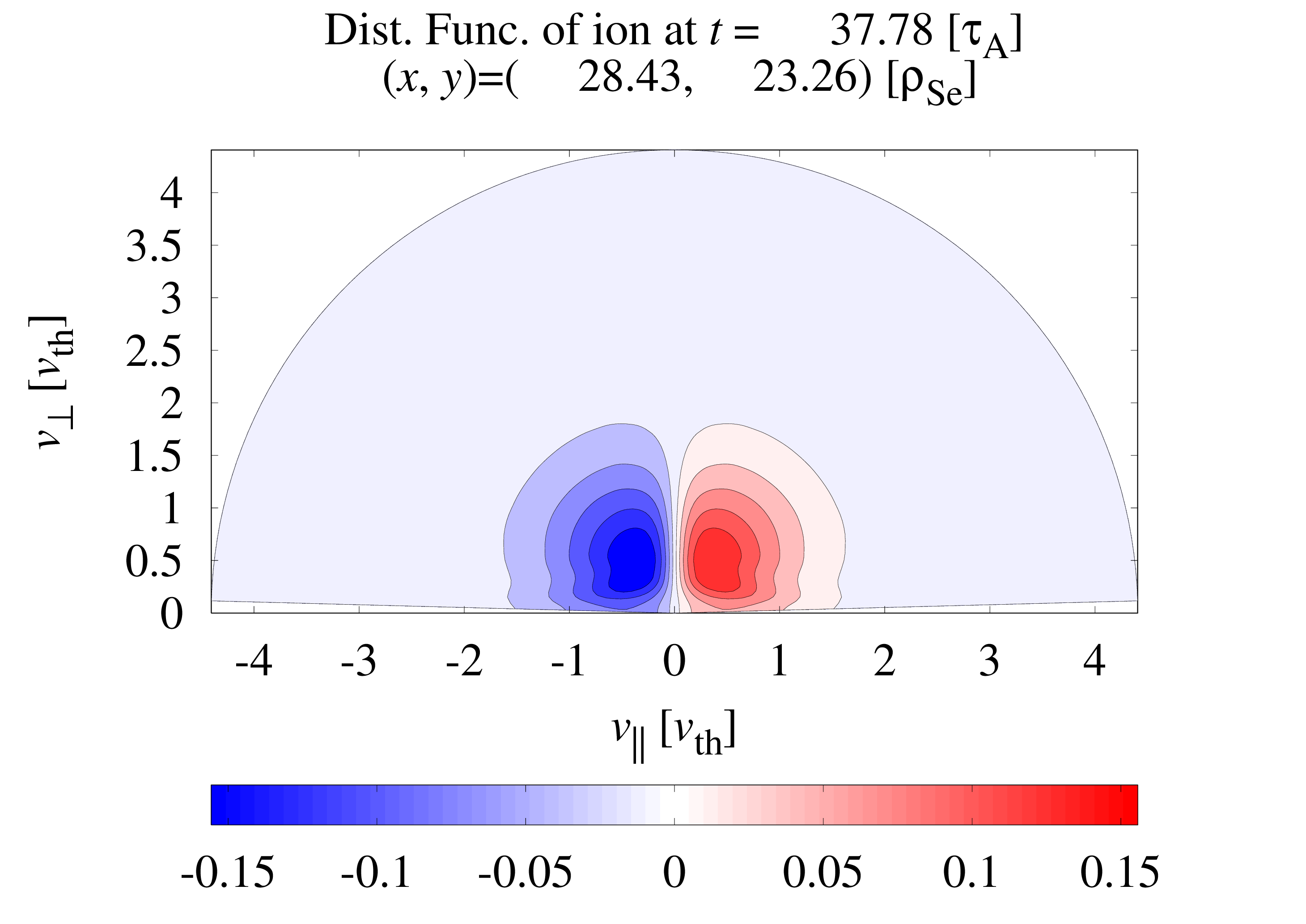}
   \includegraphics[scale=0.3]{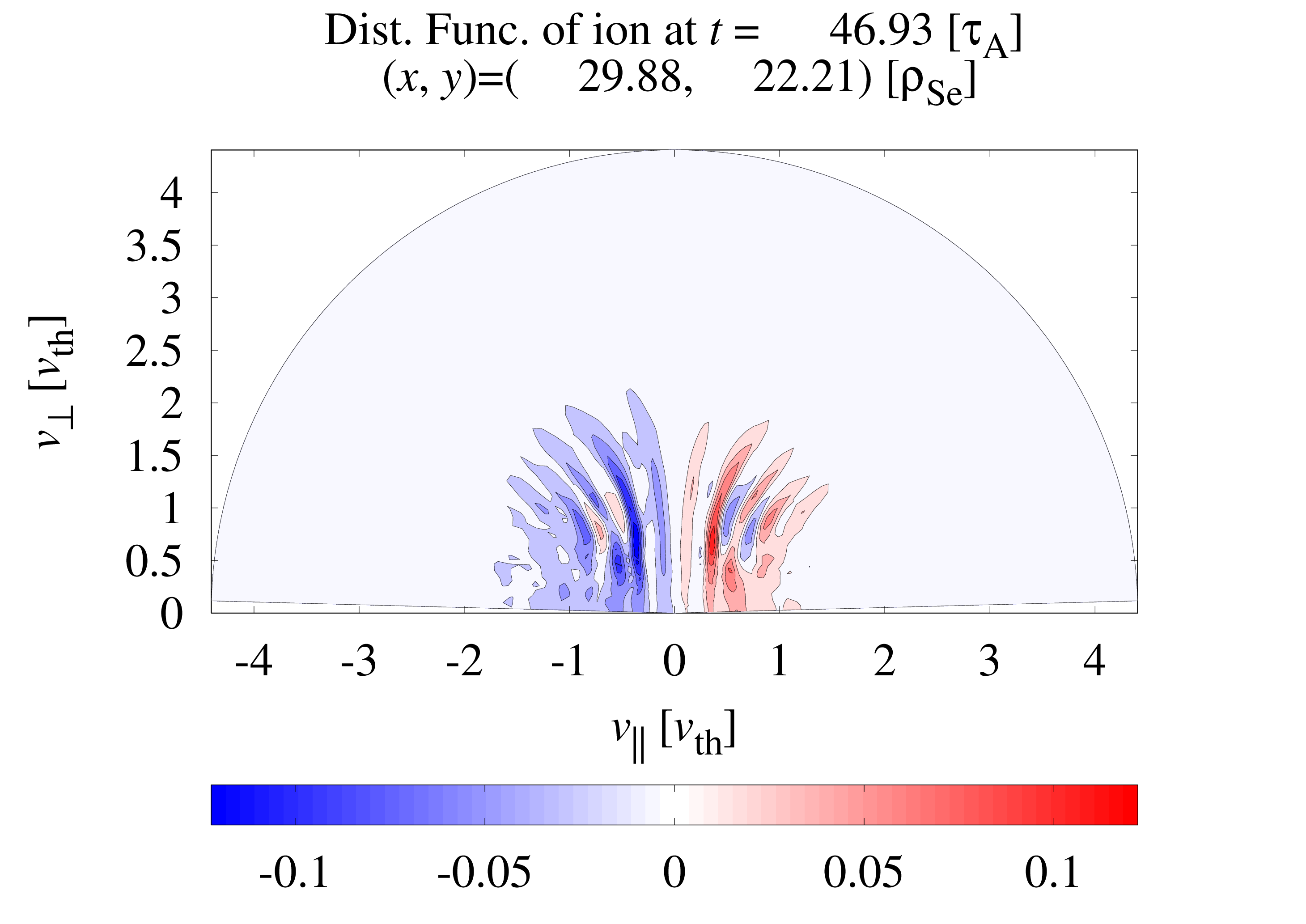}
   \caption{\label{fig:dstfn_b1_i}Velocity space structures of ion
   distribution function for $\betae=1$, normalized by
   $\varepsilon n_{0}/(\sqrt{\pi}\vthi)^{3}$.}
  \end{center}
 \end{figure}

For the $\betae=1$ case, we perform the same diagnostics for both
electrons and ions in Figs.~\ref{fig:diss_b1_e}--\ref{fig:dstfn_b1_i}.
For this case, the electron dissipation occurs, again, along the
separatrix, but confined in narrow strips.
At later times, the electron heating is restricted to
the narrow edge region of the primary island. Looking at the structure
of the electron distribution function at the point where the dissipation
is largest, we see narrow parallel phase-mixing structures. The Landau
resonance is pronounced for $v_{\parallel}<\vthe$. 
Note that $\vthe
\left(B_{\perp}/B_{z0}\right)/\VA=\sqrt{\betae/\sigma}=10$.
More electrons are, thus, in resonance at higher-$\betae$, thereby
enhancing the local strength of phase mixing as an energy dissipation
process.

At late times, the ion and electron dissipation rates become comparable.
We show spatial distributions of the ion dissipation rate at
the time of maximum reconnection rate, $t/\tauA=37.8$, and maximum of the ion
dissipation rate, $t/\tauA=47$, in Fig.~\ref{fig:diss_b1_i}. It is clearly
shown that the ion dissipation is localized to the center of the domain.
While the reconnection process proceeds, ion flows in the $z$ direction
develop as well as the $\EB$ drift in the $x-y$ plane. Collisional
damping of those flows (ion viscosity) are the main contributions to the
ion dissipation at $t/\tauA=37.8$, though this is only a small fraction
of the dissipation that occurs later as a result of phase mixing.
As shown in the distribution function plot in Fig.~\ref{fig:dstfn_b1_i},
the ion distribution function has a component proportional to
$v_{\parallel}$ driven by $E_z$,
and peaks at $v_{\perp}\sim0.5$, corresponding to the $\EB$ drift
velocity. No phase-mixing structures are visible at this time.
At later times, 
the ion dissipation is large in the secondary island at the center of
the domain. At this stage, the ion distribution function displays
significant structure both in the parallel and perpendicular velocity
directions, indicating that the overall dissipation is due to both
parallel (linear) and perpendicular (nonlinear) phase mixing.
\note{
Perpendicular phase mixing is indeed expected to be large for
$\betae\sim1$ since $k_{\perp}\rhoi\sim\rhoi/\de\gg1$. However, since
it is a nonlinear dissipation mechanism, perpendicular phase mixing can
only be significant when the amplitudes of the perturbed fields become
sufficiently large.}

\begin{figure}
  \begin{center}
   \includegraphics[scale=0.22]{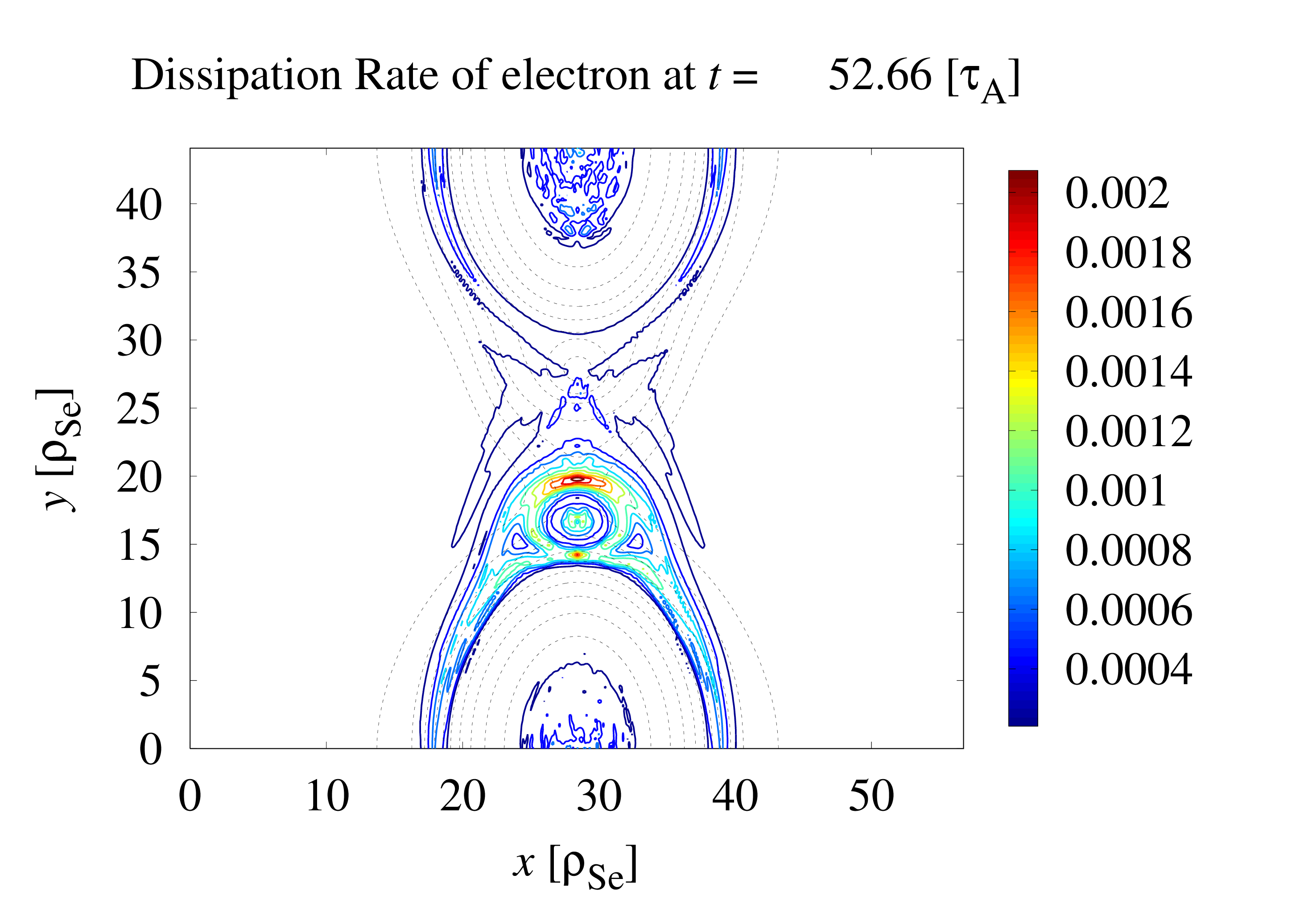}
   \includegraphics[scale=0.22]{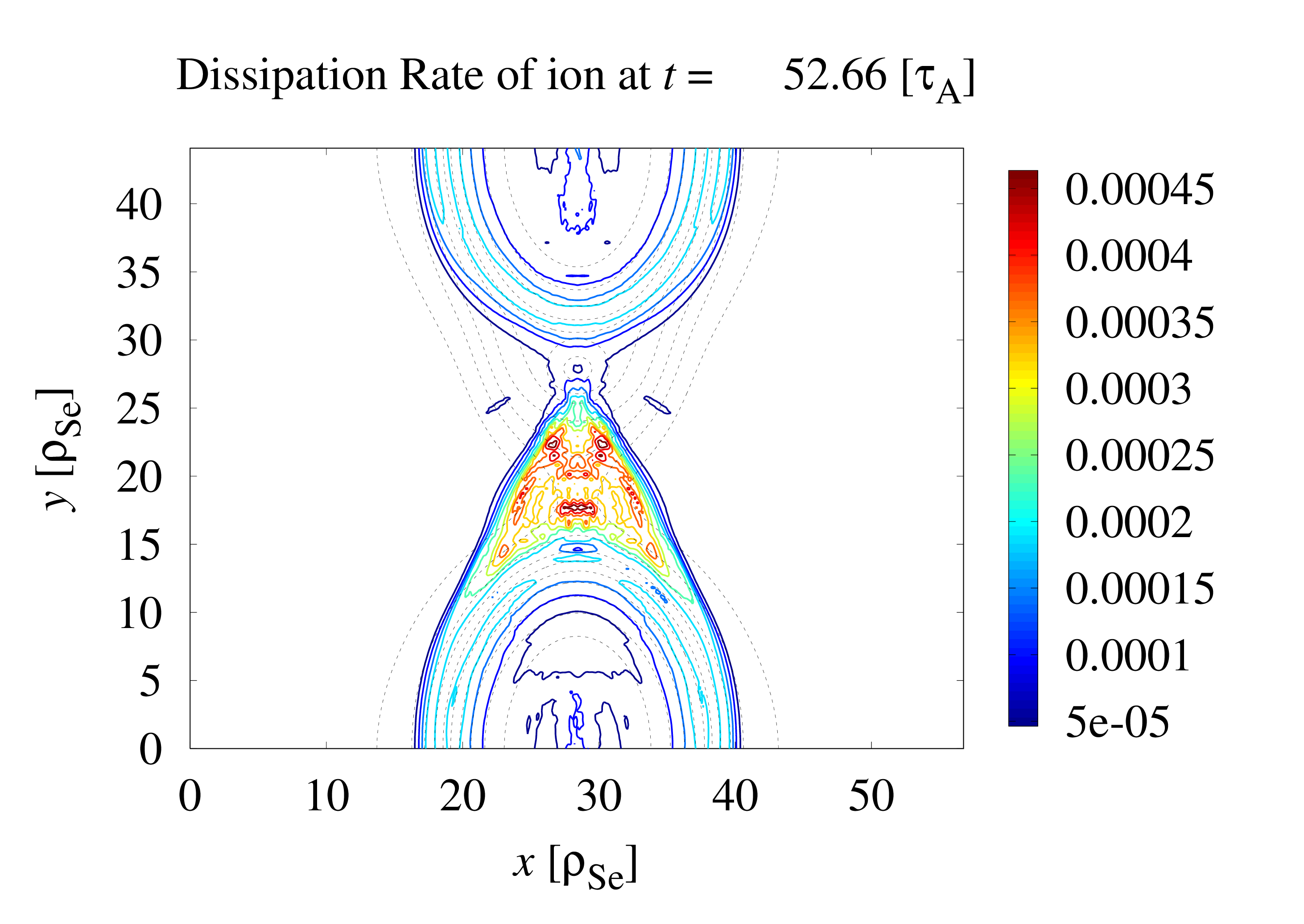}
   \caption{\label{fig:diss_b1_plasmoid}Spatial distribution of the
   dissipation rate of electrons and ions for $\betae=1$, when the
   plasmoid moves.}
  \end{center}
 \end{figure}

\revise{Around $t/\tauA=52$, the secondary island moves due to the
numerical noise, and the electron dissipation significantly increases.
\nuno{Spatial distributions of the dissipation rate of electrons and
ions at this time are shown in Fig.~\ref{fig:diss_b1_plasmoid}. We see
that the electron dissipation at this time occurs at the newly formed
$X$ point, and is stronger compared with that at earlier times,
indicating re-activation of the heating process.
The ion dissipation does not significantly increase at the plasmoid
ejection, but just spreads outside the secondary plasmoid.}}

\section{Conclusion}
\label{sec:conclusion}

We have reported gyrokinetic simulations of magnetic reconnection
in weakly collisional, \nuno{strongly magnetized} plasmas at varying
values of $\betae$.


The peak reconnection rate we find is $\sim 0.1$, as is often reported
in the weak-guide-field case, and weakly decreases with increasing
$\betae$. We also observe that, as $\betae$ increases, the reconnection
site becomes unstable to secondary island formation.

During reconnection, phase-mixing structures slowly develop.
Electron heating occurs after the dynamical reconnection process
has ceased, and continues long after.
The ion heating is comparable to the electron heating for $\betae\sim 1$,
and insignificant at lower values of $\betae$.
The fraction of energy dissipation to released magnetic energy is a
complicated function of the local strength, area, and 
duration of the phase-mixing process.

The electron heating that we measure in our simulations is caused by
parallel phase mixing. It initially occurs along the separatrix, and
slowly spreads to the interior of the primary magnetic island.
Phase mixing is most pronounced for electrons streaming along
the magnetic field at velocities $v_{\parallel}
\left(B_{\perp}/B_{z0}\right) \sim
\VA$, where $\VA$ is the perpendicular Alfv\'en velocity.
For our lowest $\betae$ case, this implies $v_{\parallel}\sim \vthe$,
whereas at larger $\betae$ the resonance condition is satisfied for
$v_{\parallel}<\vthe$. Consequently, electron dissipation is more
efficient at higher values of $\betae$.
The ion heating, on the other hand, does not spread,
\revise{as long as the plasmoid is stationary},
occurring only at the reconnection site, and later inside the secondary
island that is formed. For ions, both parallel and perpendicular
phase-mixing processes are active.
\revise{It is also found that, once the secondary island moves, the 
electron heating becomes active at the new $X$ point.}


In summary, we have shown that electron and ion bulk heating 
via phase mixing is a significant energy dissipation channel in
\nuno{strong-guide-field} 
reconnection, extending the results first reported
in \citet{LoureiroSchekochihinZocco_13} to the regime of $\betae\sim
1$.
These results, therefore, underscore the importance of retaining 
finite collisions in reconnection studies, 
even if the reconnection process itself is collisionless.

\nuno{
In addition, it is perhaps worth noting that the usual particle
energization mechanisms found in weak-guide-field reconnection (Fermi
acceleration \citep{DrakeSwisdakChe_06}) cannot be efficient for
reconnection in strongly magnetized plasmas because the curvature drift
associated with the Fermi mechanism is small. 
Our results may therefore point to an alternative heating mechanism valid
when the guide field is much larger than the reconnecting field. Scaling
studies such as those by \citet{TenbargeDaughtonKarimabadi_14} may help
identify the regions of parameter space where each of these mechanisms
(Fermi or phase mixing) predominates.
}\\


\section*{Acknowledgements}
This work was supported by JSPS KAKENHI Grant Number 24740373. This work
was carried out using the HELIOS supercomputer system at Computational
Simulation Centre of International Fusion Energy Research Centre
(IFERC-CSC), Aomori, Japan, under the Broader Approach collaboration
between Euratom and Japan, implemented by Fusion for Energy and JAEA.
N. F. L. was supported by Funda\c{c}\~ao para a Ci\^encia e a Tecnologia
through grants Pest-OE/SADG/LA0010/2011, IF/00530/2013 and PTDC/FIS/118187/2010.

\appendix
\renewcommand{\thefigure}{A.\arabic{figure}}
\setcounter{figure}{0}
\section{Velocity space convergence}
\label{sec:vconv}

\begin{figure}
 \begin{center}
  \includegraphics[scale=0.7]{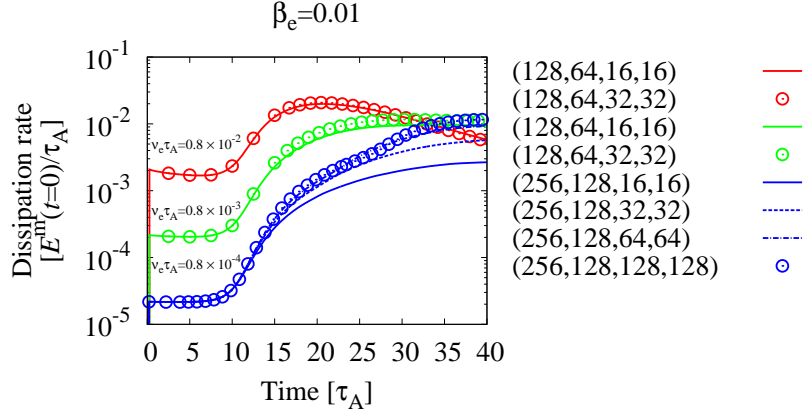}
  \vspace*{3em}
  \caption{\label{fig:vconv}Velocity-space-grid convergence for
  $\betae=0.01$ and
  $\nue\tauA=0.8\times10^{-4}\sim0.8\times10^{-2}$. The legend
  indicates the number of grid points in $x$, $y$, 
  $\lambda$, and $E$ direction, respectively: ($N_x$, $N_y$,
  $N_\lambda$, $N_{E}$).}
 \end{center}
\end{figure}

To determine the velocity-space resolution required to accurately
capture electron and ion heating in our weakly collisional simulations,
we perform a convergence test for the $\betae=0.01$ case. We run for
$\nue\tauA=0.8\times 10^{-4}, 0.8\times10^{-3}, 0.8\times10^{-2}$. Plotted
in Fig.~\ref{fig:vconv} are the dissipation rates of electrons with
several velocity-space grid sizes. In the linear phase, $t/\tauA\leq10$,
there is no phase mixing and, therefore, a relatively coarse
velocity-space grid is sufficient.
However, for the most collisionless
case $\nue\tauA=0.8\times10^{-4}$, we see that the velocity-space
resolution greatly affects the dissipation rate
in the nonlinear stage, as expected due to the development of fine
velocity scales.
For the most collisionless case, the number of velocity
grid points must be greater than $(N_\lambda, N_E)=(64, 64)$, while $(16,16)$
is sufficient for the more collisional cases.



\bibliographystyle{jpp}

\bibliography{gk_mr_heating}

\providecommand{\noopsort}[1]{}
\begin{thebibliography}{24}
\expandafter\ifx\csname natexlab\endcsname\relax\def\natexlab#1{#1}\fi

\bibitem[Abel {\em et~al.\/}(2008)Abel, Barnes, Cowley, Dorland \&
  Schekochihin]{AbelBarnesCowley_08}
{\sc Abel, I.~G., Barnes, M., Cowley, S.~C., Dorland, W. \& Schekochihin,
  A.~A.} 2008 Linearized model {Fokker-Planck} collision operators for
  gyrokinetic simulations {I.} {T}heory. {\em Phys. Plasmas\/} {\bf 15}~(12),
  122509.

\bibitem[Barnes {\em et~al.\/}(2009)Barnes, Abel, Dorland, Ernst, Hammett,
  Ricci, Rogers, Schekochihin \& Tatsuno]{BarnesAbelDorland_09}
{\sc Barnes, M., Abel, I.~G., Dorland, W., Ernst, D.~R., Hammett, G.~W., Ricci,
  P., Rogers, B.~N., Schekochihin, A.~A. \& Tatsuno, T.} 2009 Linearized model
  {Fokker-Planck} collision operators for gyrokinetic simulations {II.}
  {N}umerical implementation and tests. {\em Phys. Plasmas\/} {\bf 16}~(7),
  072107.

\bibitem[Dahlin {\em et~al.\/}(2014)Dahlin, Drake \&
  Swisdak]{DahlinDrakeSwisdak_14}
{\sc Dahlin, J.~T., Drake, J.~F. \& Swisdak, M.} 2014 The mechanisms of
  electron heating and acceleration during magnetic reconnection. {\em Phys.
  Plasmas\/} {\bf 21}~(9), 092304.

\bibitem[Dorland \& Hammett(1993)]{DorlandHammett_93}
{\sc Dorland, W. \& Hammett, G.~W.} 1993 Gyrofluid turbulence models with
  kinetic effects. {\em Phys. Fluids B\/} {\bf 5}~(3), 812--835.

\bibitem[Drake {\em et~al.\/}(2006)Drake, Swisdak, Che \&
  Shay]{DrakeSwisdakChe_06}
{\sc Drake, J.~F., Swisdak, M., Che, H. \& Shay, M.~A.} 2006 Electron
  acceleration from contracting magnetic islands during reconnection. {\em
  Nature\/} {\bf 443}~(5), 553--556.

\bibitem[Fiksel {\em et~al.\/}(2009)Fiksel, Almagri, Chapman, Mirnov, Ren,
  Sarff \& Terry]{FikselAlmagriChapman_09}
{\sc Fiksel, G., Almagri, A.~F., Chapman, B.~E., Mirnov, V.~V., Ren, Y., Sarff,
  J.~S. \& Terry, P.~W.} 2009 Mass-dependent ion heating during magnetic
  reconnection in a laboratory plasma. {\em Phys. Rev. Lett.\/} {\bf 103}~(14),
  145002.

\bibitem[Fitzpatrick(2010)]{Fitzpatrick_10}
{\sc Fitzpatrick, R.} 2010 Magnetic reconnection in weakly collisional highly
  magnetized electron-ion plasmas. {\em Phys. Plasmas\/} {\bf 17}~(4), 042101.

\bibitem[Howes {\em et~al.\/}(2006)Howes, Cowley, Dorland, Hammett, Quataert \&
  Schekochihin]{HowesCowleyDorland_06}
{\sc Howes, G.~G., Cowley, S.~C., Dorland, W., Hammett, G.~W., Quataert, E. \&
  Schekochihin, A.~A.} 2006 Astrophysical gyrokinetics: {B}asic equations and
  linear theory. {\em Astrophys. J.\/} {\bf 651}~(1), 590--614.

\bibitem[Hsu {\em et~al.\/}(2001)Hsu, Carter, Fiksel, Ji, Kulsrud \&
  Yamada]{HsuCarterFiksel_01}
{\sc Hsu, S.~C., Carter, T.~A., Fiksel, G., Ji, H., Kulsrud, R.~M. \& Yamada,
  M.} 2001 Experimental study of ion heating and acceleration during magnetic
  reconnection. {\em Phys. Plasmas\/} {\bf 8}~(5), 1916--1928.

\bibitem[Landau(1946)]{Landau_46}
{\sc Landau, L.~D.} 1946 On the vibrations of the electronic plasma. {\em Zh.
  Eksp. Teor. Fiz.\/} {\bf 16}~(7), 574--586.

\bibitem[Loureiro {\em et~al.\/}(2007)Loureiro, Schekochihin \&
  Cowley]{LoureiroSchekochihinCowley_07}
{\sc Loureiro, N.~F., Schekochihin, A.~A. \& Cowley, S.~C.} 2007 Instability of
  current sheets and formation of plasmoid chains. {\em Phys. Plasmas\/} {\bf
  14}~(10), 100703.

\bibitem[Loureiro {\em et~al.\/}(2013{\natexlab{{\em a\/}}})Loureiro,
  Schekochihin \& Uzdensky]{LoureiroSchekochihinUzdensky_13}
{\sc Loureiro, N.~F., Schekochihin, A.~A. \& Uzdensky, D.~A.}
  2013{\natexlab{{\em a\/}}} Plasmoid and {Kelvin-Helmholtz} instabilities in
  {Sweet-Parker} current sheets. {\em Phys. Rev. E\/} {\bf 87}, 013102.

\bibitem[Loureiro {\em et~al.\/}(2013{\natexlab{{\em b\/}}})Loureiro,
  Schekochihin \& Zocco]{LoureiroSchekochihinZocco_13}
{\sc Loureiro, N.~F., Schekochihin, A.~A. \& Zocco, A.} 2013{\natexlab{{\em
  b\/}}} Fast collisionless reconnection and electron heating in strongly
  magnetised plasmas. {\em Phys. Rev. Lett.\/} {\bf 111}~(2), 025002.

\bibitem[Numata {\em et~al.\/}(2011)Numata, Dorland, Howes, Loureiro, Rogers \&
  Tatsuno]{NumataDorlandHowes_11}
{\sc Numata, R., Dorland, W., Howes, G.~G., Loureiro, N.~F., Rogers, B.~N. \&
  Tatsuno, T.} 2011 Gyrokinetic simulations of the tearing instability. {\em
  Phys. Plasmas\/} {\bf 18}~(11), 112106.

\bibitem[Numata {\em et~al.\/}(2010)Numata, Howes, Tatsuno, Barnes \&
  Dorland]{NumataHowesTatsuno_10}
{\sc Numata, R., Howes, G.~G., Tatsuno, T., Barnes, M. \& Dorland, W.} 2010
  {\tt AstroGK}: {A}strophysical gryokinetics code. {\em J. Comput. Phys.\/}
  {\bf 229}~(24), 9347--9372.

\bibitem[Numata \& Loureiro(2014)]{NumataLoureiro_14}
{\sc Numata, R. \& Loureiro, N.~F.} 2014 Electron and ion heating during
  magnetic reconnection in weakly collisional plasmas. {\em JPS Conf. Proc.\/}
  {\bf 1}, 015044.

\bibitem[Ono {\em et~al.\/}(1996)Ono, Yamada, Akao, Tajima \&
  Matsumoto]{OnoYamadaAkao_96}
{\sc Ono, Y., Yamada, M., Akao, T., Tajima, T. \& Matsumoto, R.} 1996 Ion
  acceleration and direct ion heating in three-component magnetic reconnection.
  {\em Phys. Rev. Lett.\/} {\bf 76}~(18), 3328--3331.

\bibitem[Porcelli(1991)]{Porcelli_91}
{\sc Porcelli, F.} 1991 Collisionless $m=1$ tearing mode. {\em Phys. Rev.
  Lett.\/} {\bf 66}~(4), 425--428.

\bibitem[Schekochihin {\em et~al.\/}(2009)Schekochihin, Cowley, Dorland,
  Hammett, Howes, Quataert \& Tatsuno]{SchekochihinCowleyDorland_09}
{\sc Schekochihin, A.~A., Cowley, S.~C., Dorland, W., Hammett, G.~W., Howes,
  G.~G., Quataert, E. \& Tatsuno, T.} 2009 Astrophysical gyrokinetics:
  {K}inetic and fluid turbulent cascades in magnetized weakly collisional
  plasmas. {\em Astrophys. J. Suppl. Ser.\/} {\bf 182}~(1), 310--377.

\bibitem[Spitzer \& H{\"{a}}rm(1953)]{SpitzerHarm_53}
{\sc Spitzer, Jr., L. \& H{\"{a}}rm, R.} 1953 Transport phenomena in a
  completely ionized gas. {\em Phys. Rev.\/} {\bf 89}~(5), 977--981.

\bibitem[Tatsuno {\em et~al.\/}(2009)Tatsuno, Dorland, Schekochihin, Plunk,
  Barnes, Cowley \& Howes]{TatsunoDorlandSchekochihin_09}
{\sc Tatsuno, T., Dorland, W., Schekochihin, A.~A., Plunk, G.~G., Barnes, M.,
  Cowley, S.~C. \& Howes, G.~G.} 2009 Nonlinear phase mixing and phase-space
  cascade of entropy in gyrokinetic plasma turbulence. {\em Phys. Rev. Lett.\/}
  {\bf 103}~(1), 015003.

\bibitem[TenBarge {\em et~al.\/}(2014)TenBarge, Daughton, Karimabadi, Howes \&
  Dorland]{TenbargeDaughtonKarimabadi_14}
{\sc TenBarge, J.~M., Daughton, W., Karimabadi, H., Howes, G.~G. \& Dorland,
  W.} 2014 Collisionless reconnection in the large guide field regime:
  {G}yrokinetic versus particle-in-cell simulations. {\em Phys. Plasmas\/} {\bf
  21}~(2), 020708.

\bibitem[Yamada {\em et~al.\/}(2010)Yamada, Kulsrud \& Ji]{YamadaKulsrudJi_10}
{\sc Yamada, M., Kulsrud, R. \& Ji, H.} 2010 Magnetic reconnection. {\em Rev.
  Mod. Phys.\/} {\bf 82}~(1), 603--664.

\bibitem[Zocco \& Schekochihin(2011)]{ZoccoSchekochihin_11}
{\sc Zocco, A. \& Schekochihin, A.~A.} 2011 Reduced fluid-kinetic equations for
  low-frequency dynamics, magnetic reconnection and electron heating in
  low-beta plasmas. {\em Phys. Plasmas\/} {\bf 18}~(10), 102309.

\end{thebibliography}

\end{document}